\documentclass[english]{article}
\usepackage[T1]{fontenc}
\usepackage[latin9]{inputenc}
\usepackage{float}
\usepackage{graphicx}
\usepackage{esint}
\PassOptionsToPackage{normalem}{ulem}
\usepackage{ulem}

\makeatletter
\newcommand{\lyxaddress}[1]{
	\par {\raggedright #1
	\vspace{1.4em}
	\noindent\par}
}

\usepackage{babel}
\usepackage[superscript,biblabel]{cite}

\makeatother

\usepackage{babel}
\begin{document}
\title{Aqueous-alcohol mixtures in dimension two: miscibility and micro-segregation}
\author{Camille de la Vaissiere, Ayse Butuner and Aurélien Perera\thanks{Corresponding author:aurelien.perera@sorbonne-universite.fr}}
\maketitle

\lyxaddress{Laboratoire de Physique Théorique de la Matière Condensée (UMR CNRS
7600), Sorbonne Université, 4 Place Jussieu, F75252, Paris cedex 05,
France.}
\begin{abstract}
Two dimensional site interaction models of water and alcohols are
mixed in various proportions and studied by Monte Carlo simulations,
with the purpose to clarify problems related to simulation of real
micro-heterogeneous systems. Three alcohols are considered, methanol,
pentanol and octanol. The main finding is that, while real alcohols
demix with water from butanol onward, their 2D analogs are always
fully miscible, while developing increasingly pronounced micro-segregation
as the alcohol tail length increases. This is not a consequence of
the intrinsically higher fluctuations in 2D, but rather a reorganization
of these fluctuations under the charge ordering mechanism. The second
finding is that water drives the micro-segregation through strong
self-aggregation, but this is not enough to achieve full phase separation
because of the water-alcohol contact at the outer rim of the water
domains. In this work we examine how this local heterogeneity develops
with increasing alcohol alkyl tails, monitored with the study of pair
correlation functions, structure factors and Kirkwood-Buff integrals.
The absence of clear \emph{local} self-averaging of the latter provides
an illustration of the tension between energy driven maintaining of
local structures and entropy driven global homogeneity. In that, the
2D modelisation of real hydrogen bonding mixtures allows to better
capture and reveal the physics behind the chemistry of these liquids.
\end{abstract}

\section{Introduction}

In the past two decades, many publications have put forward the micro-heterogeneous
nature of several types of liquids, ranging from aqueous mixtures
\cite{SoperNature,smith-TFE,nico-TBA,Friedrich2025,myIUPAC} to room
temperature ionic liquids \cite{margulis,Annapureddy2010,Russina2014,Triolo2007}
and how this property translates in the x-ray scattering experiments,
and particularly in relation to the underlying structure factors and
pair correlation functions\cite{Annapureddy2010,Siqueira2011,Perera2022}.
Indeed, from statistical mechanics, these quantities reflect the spatial
extent of the fluctuations, and point towards whether micro-heterogeneity
is a special type of density/concentration fluctuation, or an independent
property, and, if so, what are the signatures that allow to tell this
difference. This question has also been raised in the experimental
context, for instance in the existence of unexpected large heterogeneity
in the scattering of mixtures such as aqueous tert-butanol \cite{anisimov2011,Sedlak2014}
or aqueous tetrahydrofuran (THF) mixtures\cite{THF-problem}. This
problem has been further emphasized by the difficulties met in the
simulations of several aqueous mixtures, some of which left with no
solution, such as the spurious demixing of aqueous-tbutanol mixture
at around alcohol mole fraction x=0.1 \cite{Overduin2017,Overduin2019},
or the LCST part of the aqueous THF mixtures\cite{Brovchenko2001}.
In relation to these problems, it is interesting to note that most
systems with strong heterogeneity cannot be solved with conventional
integral equation techniques\cite{aup_faraday,kezic-Coresoft-2015},
and the origin of this particular problem has been suggested to lie
in the need for appropriate bridge functions\cite{aup_baptista}.
The connection between the bridge function and the problem of the
heterogeneity and subsequent domain formation has been investigated
recently\cite{aup-POF}.

In this context, it therefore appears as worthwhile to re-investigate
these issues with two-dimensional analogs of these systems, provided
of course that reasonable representation of the molecules could be
modeled. Two-dimensional version of water, the Mercedes-Benz (MB)
water\cite{BenNaim-MB,DillMB1998,urbicMB2000,urbicMB2002} has been
introduced in order to understand the Hbonding properties of water\cite{DillReview}.
In recent works, we have introduced a site interaction equivalent
of 2D water\cite{aupSS-water} , which produced pair correlation function
and structure factor in better agreement with real 3D water than for
the MB model. Similarly, 2D site interaction alcohol models have been
introduced \cite{AUP-2Dalc} as their MB counterpart\cite{alcTomaz}.
These models capture quite well the micro-structure of the real alcohols
as obtained through the study of the site-site pair correlation functions
and corresponding structure factor. In that, both the water and alcohol
2D models are interesting to investigate the micro-structure of the
mixture, which poses statistical problems in 3D simulations.

The important point here is that the immiscibility of water and long
alcohols is usually explained by the dominance of the repulsion between
water and the alkyl tails. Since particles with opposing interactions
equally demix in 2D, one is supposed to expect that the same repulsion
could occur in 2D and consequently lead to demixing. This is not what
that is observed, which suggests that, in three dimensions, the self
Hbonding property plays at least a role comparable to the water-alkyl
repulsion.

Beyond their intrinsic interest, such 2D models provide a simplified
environment where domain correlations can be identified more clearly
than in realistic 3D systems, thereby allowing a more direct investigation
of the relationship between micro-segregation, concentration fluctuations
and Kirkwood-Buff integrals.

This work is divided into two parts. The first part focuses on system
specific and cross system comparisons for structural properties, similarly
to studies usually conducted in most publications. The second part
focuses on problems posed by fluctuations and in analogy with the
real 3D systems, which is tackled through the analysis of the Kirkwood-Buff
integral.

\section{Models and simulation details}

\subsection{Models}

\begin{figure}[H]
\centering
\includegraphics[scale=0.3]{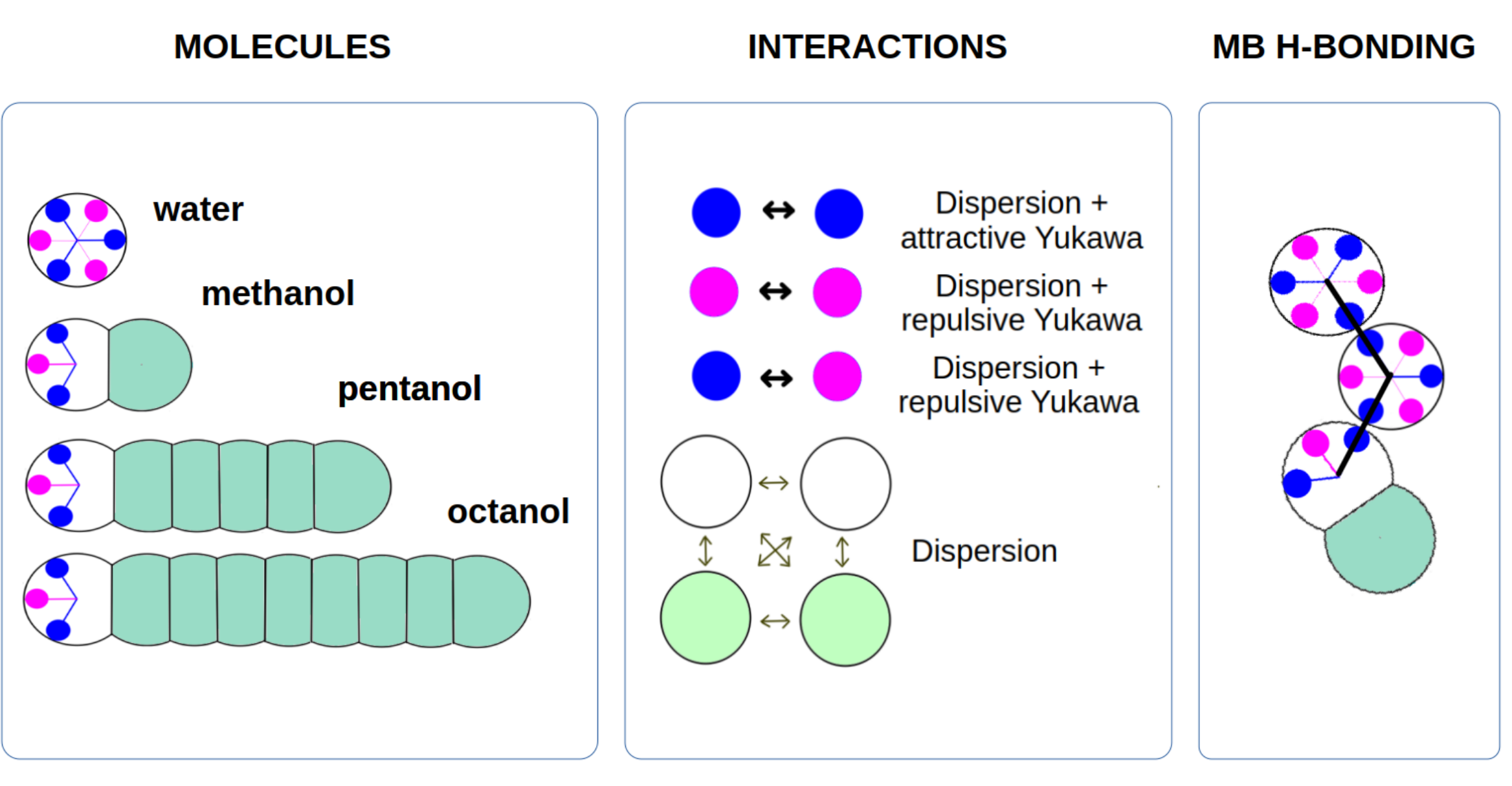}

\caption{Representation of the water and alcohol models (left), the site interactions
(middle) and Mercedes-Benz type hydrogen bonding between the central
water molecules and adjacent water and methanol molecules (right)
as shown by the thick black lines.}

\label{models}
\end{figure}

Instead of the original orientation-based Mercedes-Benz (MB) model\cite{BenNaim-MB,DillMB1998},
we employ the site-interaction versions of water and alcohol molecules
introduced previously for 2D water\cite{aupSS-water} and 2D alcohols\cite{AUP-2Dalc},
as depicted in Fig.\ref{models}. These models retain the characteristic
hydrogen-bonding pattern of the MB approach while replacing orientational
bonding rules by explicit site-site interactions. The base MB bonding
pattern based on site interaction is illustrated in the panel on the
right.

The site-site interactions between two atoms $a_{i}$ and $b_{j}$
, belonging respectively to species $a$ and $b$, are the sum of
a dispersive term and a Coulomb-like term as follows:

\begin{equation}
v_{a_{i}b_{j}}(r)=4\epsilon_{a_{i}b_{j}}\left(\frac{\sigma_{a_{i}b_{j}}}{r}\right)^{12}+\alpha_{a_{i}b_{j}}\frac{Z_{a_{i}}Z_{b_{j}}}{r}\exp\left[-\left(\frac{r-R_{a_{i}b_{j}}}{\kappa_{a_{i}b_{j}}}\right)^{2}\right]\label{V12}
\end{equation}
The cross interaction parameters $\sigma_{a_{i}b_{j}}$ and $\epsilon_{a_{i}b_{j}}$
are obtained from the Lorentz and Berthelot combining rules, respectively..
The individual parameters are listed in the Supplemental Information
document.

\subsection{Simulation details}

The packing fraction $\eta=0.6$ is chosen such that pure water is
in the liquid phase for this model. The temperature $T=2$ is chosen
since it correspond to liquid state for both water\cite{aupSS-water}
and alcohols\cite{AUP-2Dalc}.

As in Ref.{[}\#AUP-2Dalc{]} for the neat alcohols, constant NVT Monte
Carlo simulations have been performed, with a total number of molecules
around $N=1000$, depending of the initial packing conditions. In
some cases, systems of $N=2000$ molecules were simulated to test
system size effects, as discussed in Section \ref{sec:Domain-correlations-and}.
Typical solute mole fractions $x=0.2$, $x=0.5$ and $x=0.8$ have
been studied, corresponding to a range between rich solvent and rich
solute regimes. Each system was melted at high temperature $T=15$
for about 100 thousand steps, before brought by successive steps to
T=8. Then equilibration runs of 100 thousand Monte Carlo steps have
been performed, each such step consisting of trial moves of all the
N molecules. Then, between 3 and 5 indendent production runs of 150
thousand steps have been performed from decorrelated configurations.
Such configurations were constructed by heating to $T=15$ the charge-less
version of a previous run. It was necessary to do so in order to demonstrate
that the long range domain oscillation in the tail of the site-site
pair correlation functions $g_{a_{i}b_{j}}(r)$ did not satisfy self-averaging
within the simulated window, as discussed section \ref{sec:Discussion}.
This is a central point of this study, which is easier to study in
2D model systems than in 3D realistic systems, which we reported in
several previous works Refs.{[}\#Perera2017a, \#Perera2022, \#myIUPAC{]},
although it was difficult to appreciate their statistical nature properly.
Self-averaging is an important property of equilibrium statistical
mechanics\cite{self-av-BinderStauffer1997}, and it is shown to be
violated in system which have several minima such as glasses for instance\cite{self-av-WisemanDomany1995,self-av-AharonyHarris1996}.
We do not expect the systems we study to have any analogy with glasses;
.yet the presence of what could be termed \emph{local} self-averaging,
namely self-averaging over molecular scales but not necessarily over
larger domain scales, is an intriguing phenomenon, as related to the
formation of micro-heterogeneity, hence related to the physics of
fluctuations\cite{chaikin2000principles}.

In addition, a problem noticed in our previous study of pure alcohol
is enhanced in the present case of mixtures: it concerns the ringing
effect in the low $k$ part if the structure factors. This ringing
artifact arises from the conjugated effects of the statistical noise
in the tails of the $g_{a_{i}b_{j}}(r)$, the requirement to fit these
functions to a log scale before performing the 2D Fourier-Talman transform,
and the long range domain oscillations of the tails associated to
the domain correlations. Their smaller magnitude in the case of neat
alcohol is explained by the absence of the domain oscillations. Thus,
the amplified ringing effects are essentially coming from the domain
oscillations, which themselves are non-self averaging. The persistence
of these oscillations complicates the statistical convergence of long-range
observables..

\section{Results}

In this section, we first present a visual analysis of the micro-structure
of these mixtures for three typical alcohol mole fraction covering
the low value with 20\%, equimolar 50\% and finally high value with
80\%. Since these systems are strongly inhomogeneous, any snapshots
shows a typical example of how molecular species gather. In a second
subsection we present the pair correlation functions between pairs
of sites, and typically those between the charged sites. These observables
(from a simulation point of view) provide a theoretical alternative
to the snapshots, and it is interesting to find the signature of the
visually seen specificities. Finally, in a third subsection, we present
the structure factors, which are the Fourier transforms of the pair
correlation functions, and allow to enhance the medium to long range
parts of the correlations, hence providing a complementary view to
the correlations function analysis.

\subsection{Snapshots}

\begin{figure}[H]
\centering
\includegraphics[scale=0.3]{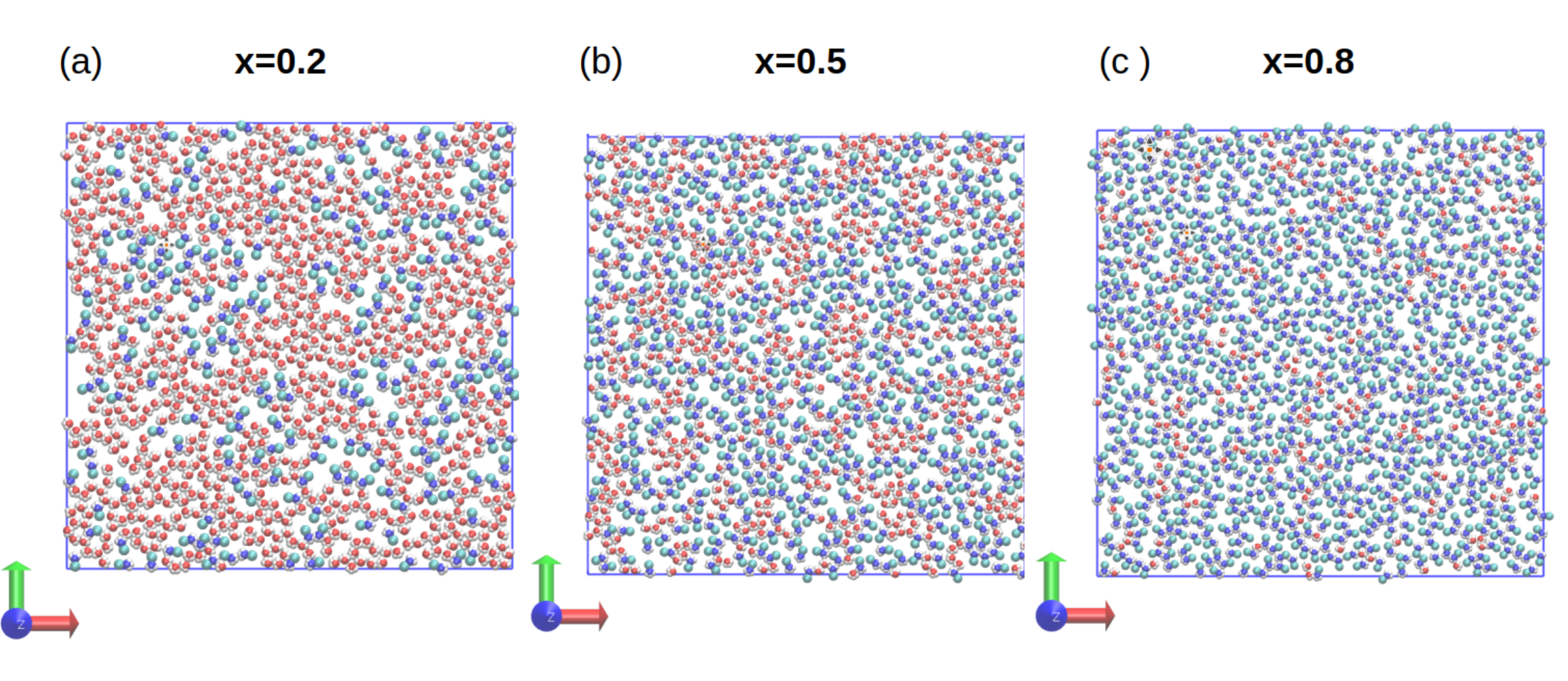}

\caption{Snapshot of the 2D aqueous methanol mixtures for concentrations $x=0.2$
(left panel), $x=0.5$ (middle panel) and $x=0.8$ (right panel).
The water and alcohol molecules are represented as in Fig.\ref{models}.}

\label{snapMW}
\end{figure}

\begin{figure}[H]
\centering
\includegraphics[scale=0.3]{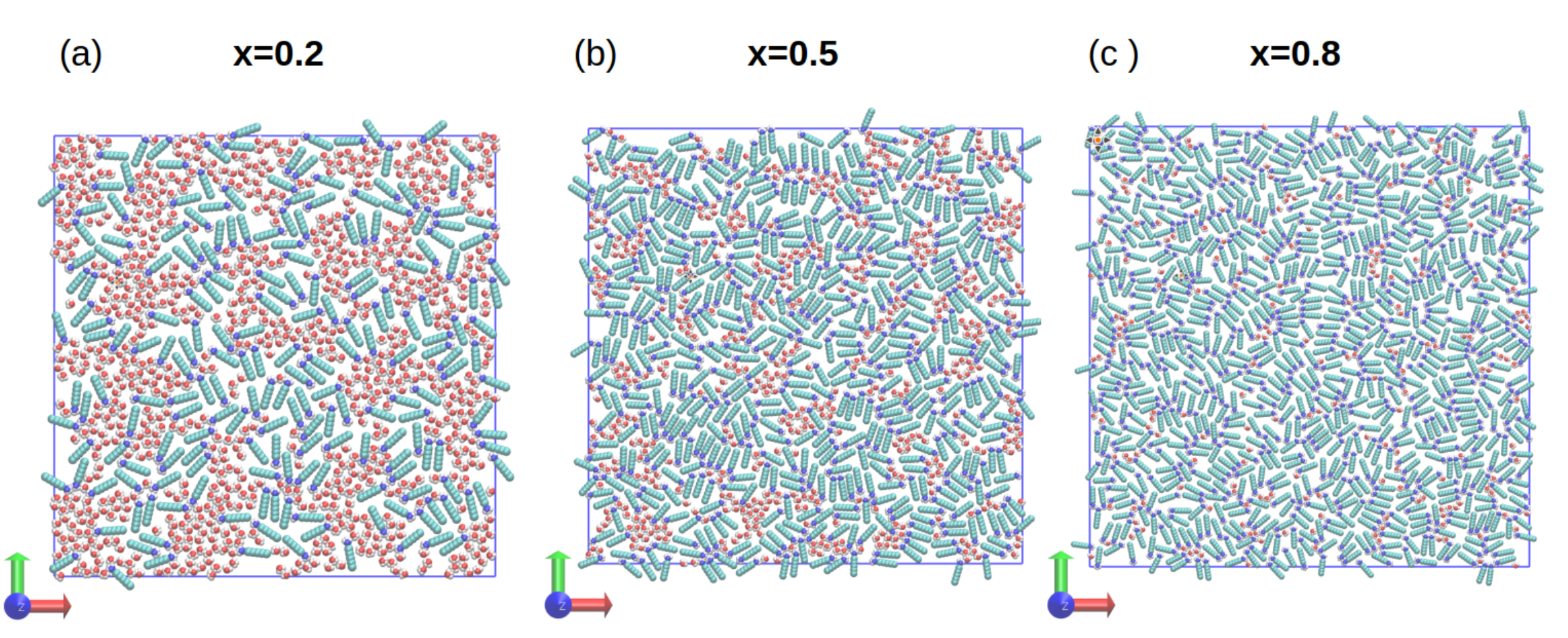}

\caption{Snapshot of the 2D aqueous pentanol mixtures for concentrations x=0.2,
x=0.5 and x=0.8. The water and alcohol molecules are represented as
in Fig.\ref{models}.}

\label{snapWP}
\end{figure}

\begin{figure}[H]
\centering
\includegraphics[scale=0.3]{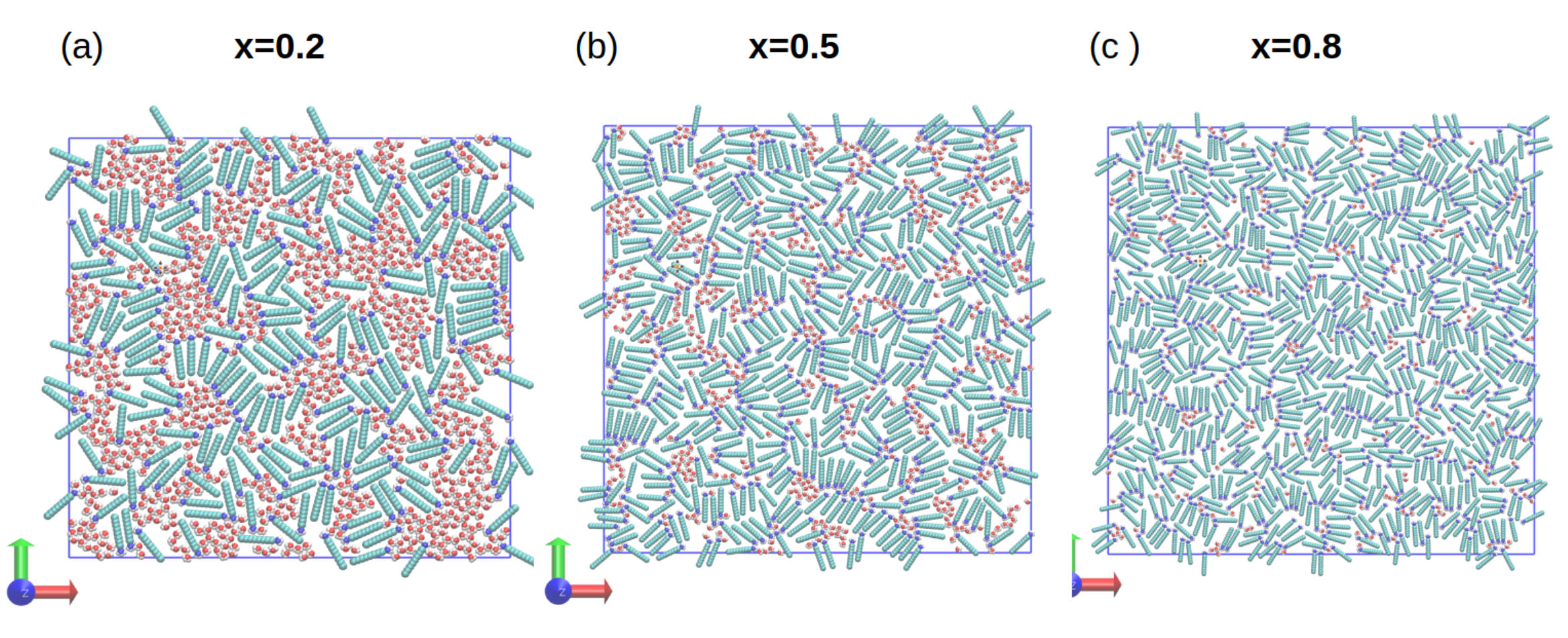}

\caption{Snapshot of the 2D aqueous octanol mixtures for concentrations x=0.2,
x=0.5 and x=0.8. The water and alcohol molecules are represented as
in Fig.\ref{models}.}

\label{snapWO}
\end{figure}

The snapshots represent the characteristic micro-structure of these
model mixtures and are detailed in Figs.\ref{snapMW}-\ref{snapWO}
below for each of the aqueous alcohol mixtures considered herein.
What can be clearly observed in all snapshots is that water tends
to group into small domains. This is more obvious for pentanol and
octanol mixtures. The case of water-methanol mixture is quite similar
to that observed in the simulations of the 3D systems\cite{SoperNature},
with a rather moderate segregation between species. 

When water is majority species, it tends to form ring-like clusters,
not always with 3 neighbours as in the 2D hexagonal crystalline structures.
When temperature is decreased toward that of liquid 2D water, water
tends to become amorphous, since small water domains have a better
tendency to Hbond.

The most striking observation is the absence of macroscopic phase
separation, as seen in Fig.\ref{snapWP} for water-pentanol, despite
the fact that real water-pentanol mixtures are only partially miscible.At
large solute concentrations, water seems to fill gaps left by the
rod-like pentanols. The study of the pair correlation in the next
section will nuance this statement.

Fig.\ref{snapWO} for aqueous octanol mixtures shows again an even
more striking feature, the clear absence of demixing at all concentrations,
when the real water-octanol mixtures show demixing over a large range
of concentrations\cite{AqOct-siepmann}. At low octanol concentration,
it can be seen that the 2D octanol molecules tend to pack in parallel
configurations, with water filling the gaps in between the packed
alkyl tail domains. We observe again that water tend to form ring-like
Hbonded domains.

The unexpected water-pentanol and water-octanol mixing could be explained
by the tendency of water to form self clustered domains. However,
it does not explain why this tendency leads to partial mixing instead
to full demixing. To verify that this behavior was not an artifact
of the preparation protocol, additional simulations were initiated
from artificially demixed configurations. In all cases the systems
evolved toward partially mixed states, indicating that the observed
micro-segregated structures correspond to thermodynamically stable
configurations rather than metastable remnants of the initial state.
It would seem that water molecules diffuse within the interstitial
regions between packed alcohol tails and subsequently reorganize into
hydrogen-bonded domains.

\subsection{Pair correlation functions}

\subsubsection{Motivation}

If each snapshot is an instantaneous sampling of the density $\rho_{i_{a}}(\mathbf{r})$
of atoms i of species a, then the site-site pair correlation function
\[
g_{i_{a}j_{b}}(r=|\mathbf{r}-\mathbf{r}'|)=\frac{<\rho_{i_{a}}(\mathbf{r})\rho_{j_{b}}(\mathbf{r})>}{\bar{\rho}_{a}\bar{\rho}_{b}}
\]
 represents the statistical average of correlation between such instantaneous
densities for two different atoms (where $\bar{\rho}_{a}$ is the
density of species $a$ - considering that each atom is unique). These
functions contains information on the atomic packings weighted by
the nature of the interactions between the two atoms, hence provide
an alternative observable to the snapshots. This goes from the distortions
seen at short range to the segregated species domain correlations
at long range. These properties differ from different types of atoms,
whether they are charged or not.

As discussed and illustrated in Ref.\cite{AUP-2Dalc} for the case
of pure 2D alcohols, due to the very different magnitude in the peak
of the $g_{a_{i}b_{j}}(r)$ functions, as well as their long rang
extent, it is sometimes convenient to plot them in either semi-log
or log-log scales, a convention that is systematically used in this
subsection. Below, we focus mostly on the correlations between the
charged atoms. Those with the tail uncharged atoms are show in the
SI document in Section 2-SI, together with other pair correlation
functions.

\subsubsection{Charge order is mostly ruled by water}

\begin{figure}[H]
\centering
\includegraphics[scale=0.3]{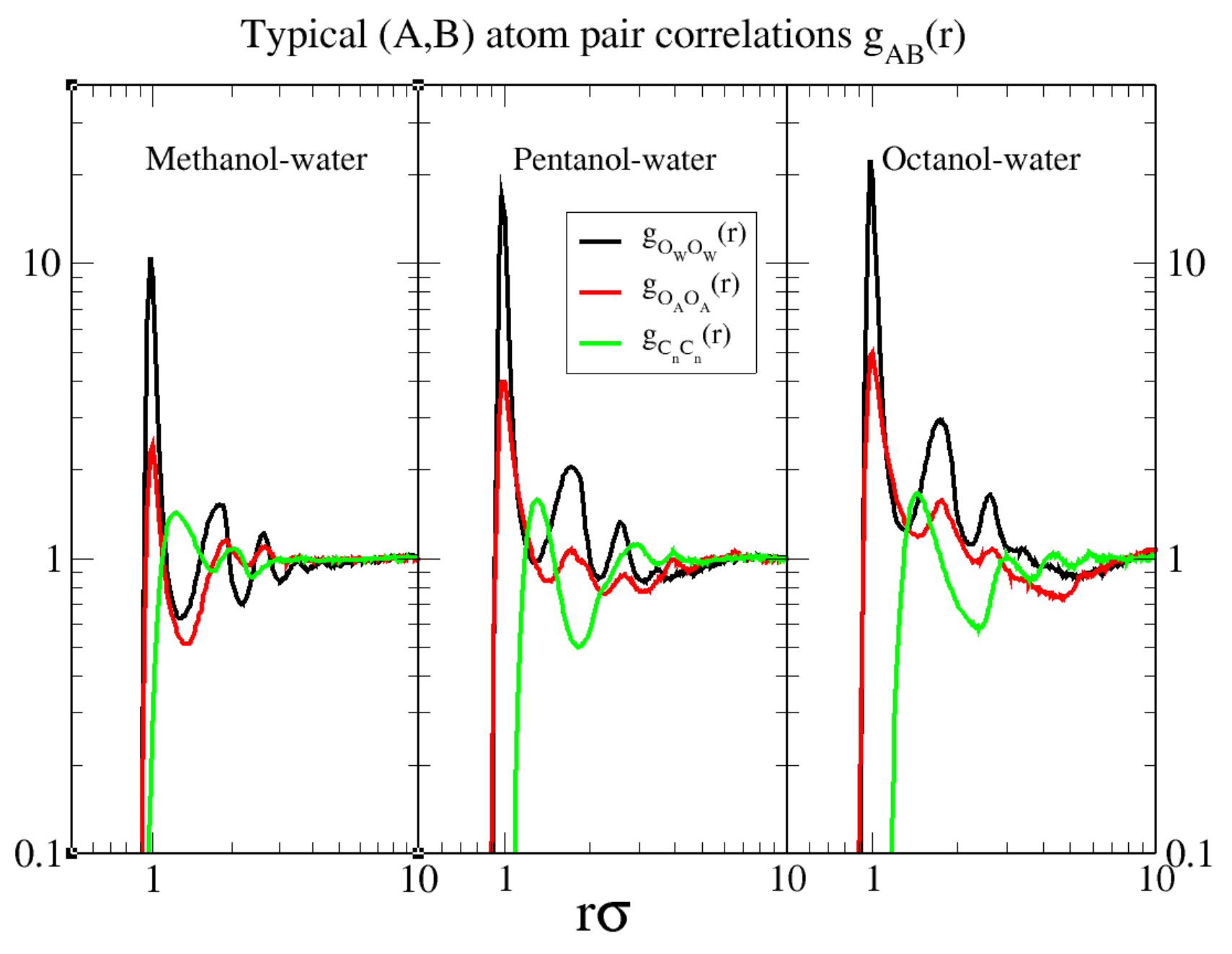}

\label{Fig=000020GR=000020COMP}

\caption{Typical log-log scale site-site correlation functions $g_{a_{i}b_{j}}(r)$
of the 3 equimolar aqueous-alcohol mixtures, namely water and alcohol
oxygen-oxygen correlations $g_{O_{W}O_{W}}$(r), $g_{O_{A}O_{A}}(r)$
, and selected neutral carbon atom correlations $g_{C_{n}C_{n}}(r)$,
where $n=1$,$3$ and $5$, for methanol, pentanol and octanol, respectively.}
\end{figure}

Fig.\ref{Fig=000020GR=000020COMP} shows a comparison between the
typical pair correlations in the mixtures, which are the dominant
charged group water-water correlations in $g_{O_{W}O_{W}}(r)$, the
corresponding alcohol correlations in $g_{O_{A}O_{A}}(r)$ and neutral
tail group correlations labeled $g_{C_{n}C_{n}}(r)$, where $n=1$
for methanol representing the methyl group, $n=3$ and $5$ for pentanol
and octanol, respectively, representing correlations between the central
methylene groups.

These correlations show what is apparent in the snapshots, that they
are mostly dominated by the water-water correlation (in black), as
an effect of the higher charges (see Table SI-I in the SI document).
The alcohol hydroxyl group correlations (in red) are much weaker than
those in pure alcohols as shown in Ref.\cite{AUP-2Dalc}, because
they are most of them are shared with water. The neutral methyl sites
have much weaker correlations in amplitude but also in packing, as
can be seen through the broader first peaks. Comparing these these
mixtures, we observe an interesting difference in the charged group
correlations. Water-methanol water-water and alcohol-alcohol correlations
show a slight depletion, seen through the small but noticeable increase
towards the asymptote 1: this is usually interpreted as chaining of
the charged groups. In the water-octanol mixture, we observe an initial
opposite behaviour stretching over the first neighbours, indicating
that both alcohol and water tend to form globular clusters. The correlations
then tend to raise, because of the domain oscillations that occur
at longer ranges. The water-pentanol correlations show behaviour similar
to that of water-octanol, but with a smaller initial clustering range.
In contrast to this observation, the correlation between the carbon
groups seem to always increase towards 1, as in a depletion mode.
This is in fact a signature that the alkyl tails tend to pack in short
parallel stacks, as seen in the snapshots of the previous section,
most visible for pentanol and octanol in the high packing fractions.
These observations are compatible with a picture where it is water
which controls the micro-structure of the mixtures. The correlations
range extends appreciably when going from methanol to octanol. 

\subsubsection{Polar head correlations}

\begin{figure}[H]
\centering
\includegraphics[scale=0.2]{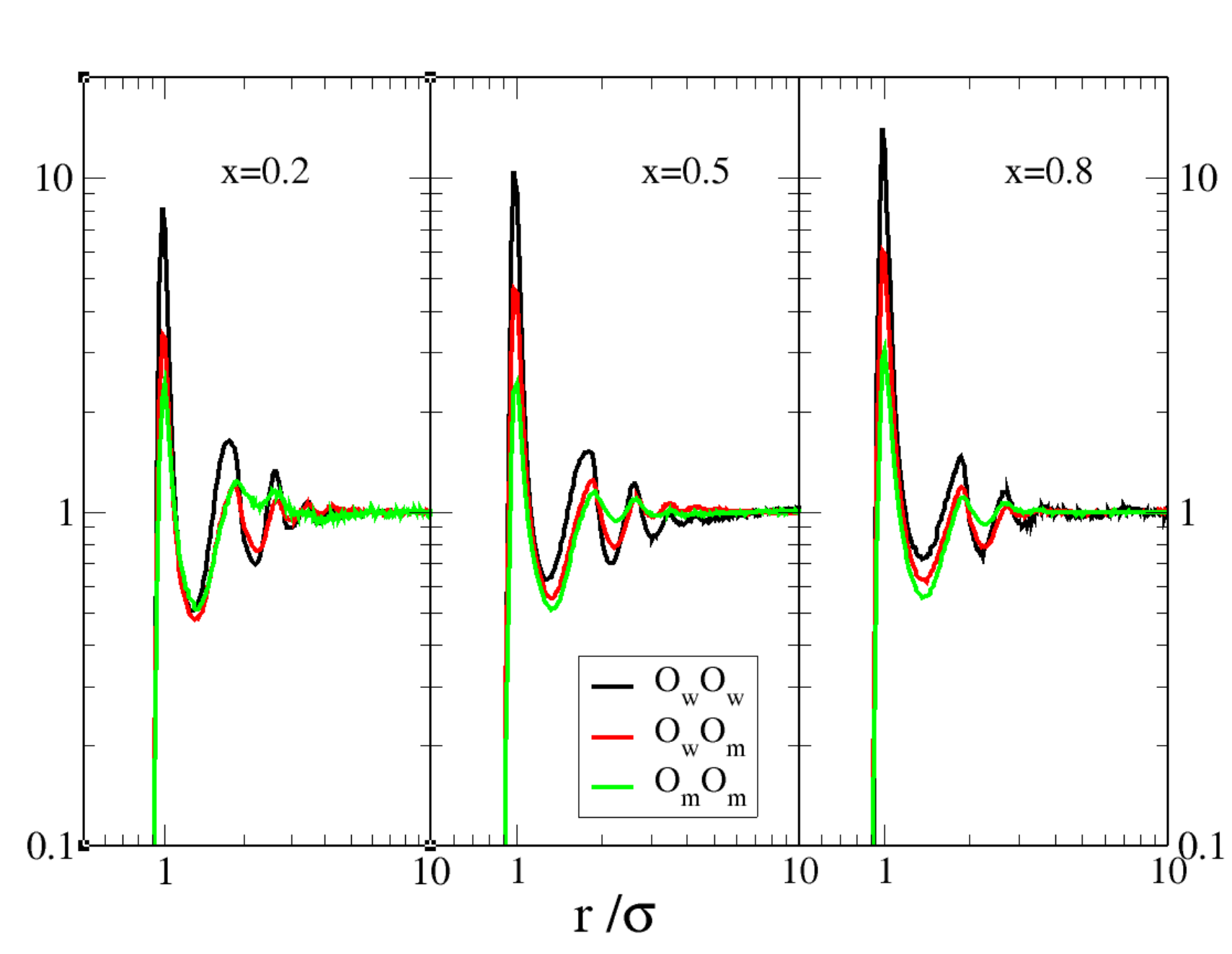}

\caption{Different types of oxygen-oxygen pair correlation functions for three
different methanol concentrations: 20\%(left panel), 50\% (middle
panel) and 80\%(right panel). The water-water correlations $g_{O_{w}O_{w}}(r)$
are show in black lines, the water-methanol cross correlations $g_{O_{w}O_{m}}(r)$
shown in red lines, and the methanol-methanol correlations $g_{O_{m}O_{m}}(r)$
shown in green lines. The symbols $O_{w}$ and $O_{m}$ stand for
oxygen atoms of \uline{w}ater and \uline{m}ethanol, respectively.}

\label{Fig=000020GR=000020WM}
\end{figure}

\begin{figure}[H]
\centering
\includegraphics[scale=0.2]{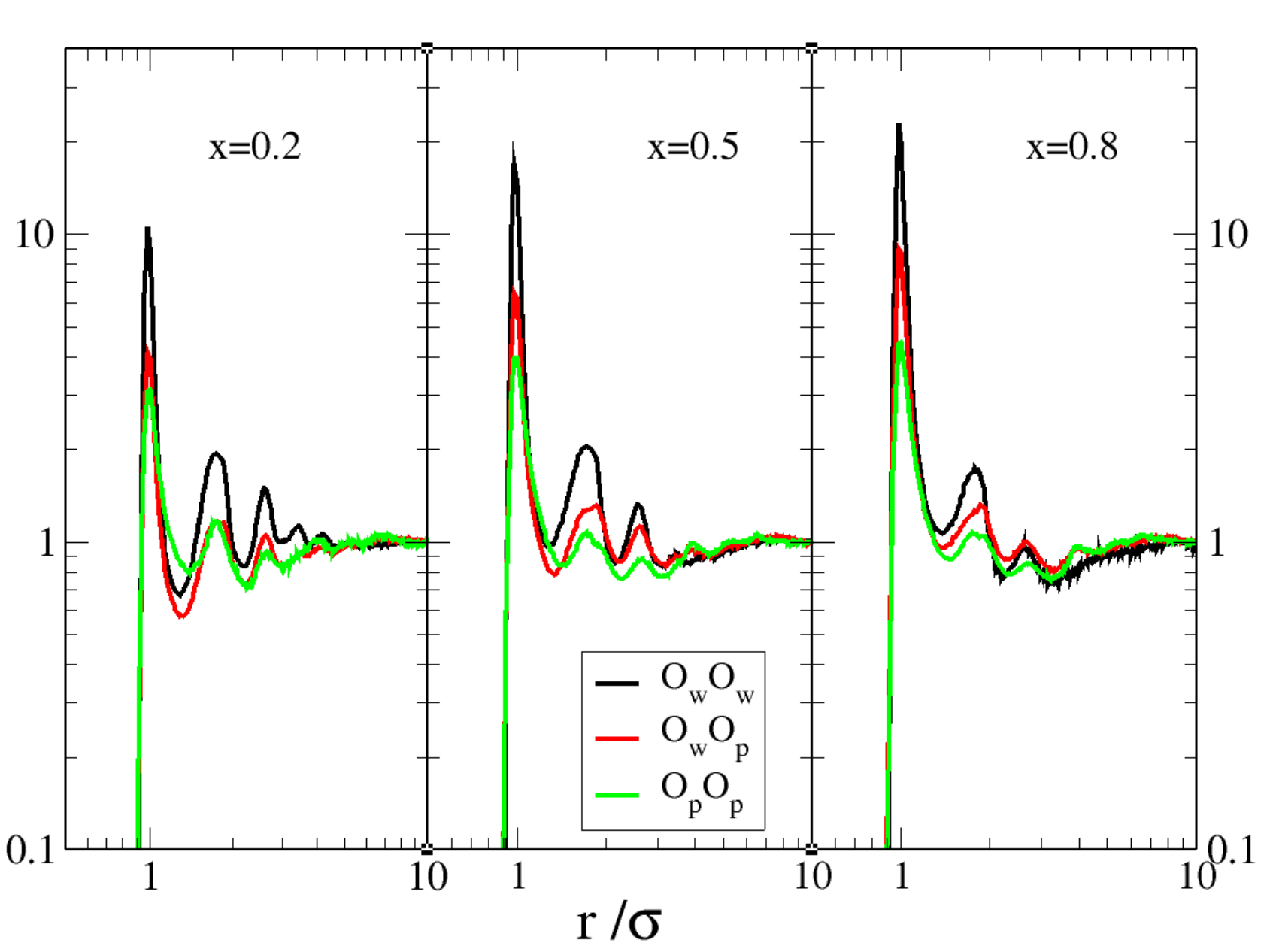}

\caption{Oxygen-oxygen pair correlation functions for three different pentanol
concentrations, with same conventions as those in Fig.\ref{Fig=000020GR=000020WM}.
The symbol $O_{p}$ stand for oxygen atoms of \uline{p}entanol.}

\label{Fig=000020GR=000020WP}
\end{figure}

\begin{figure}[H]
\centering
\includegraphics[scale=0.2]{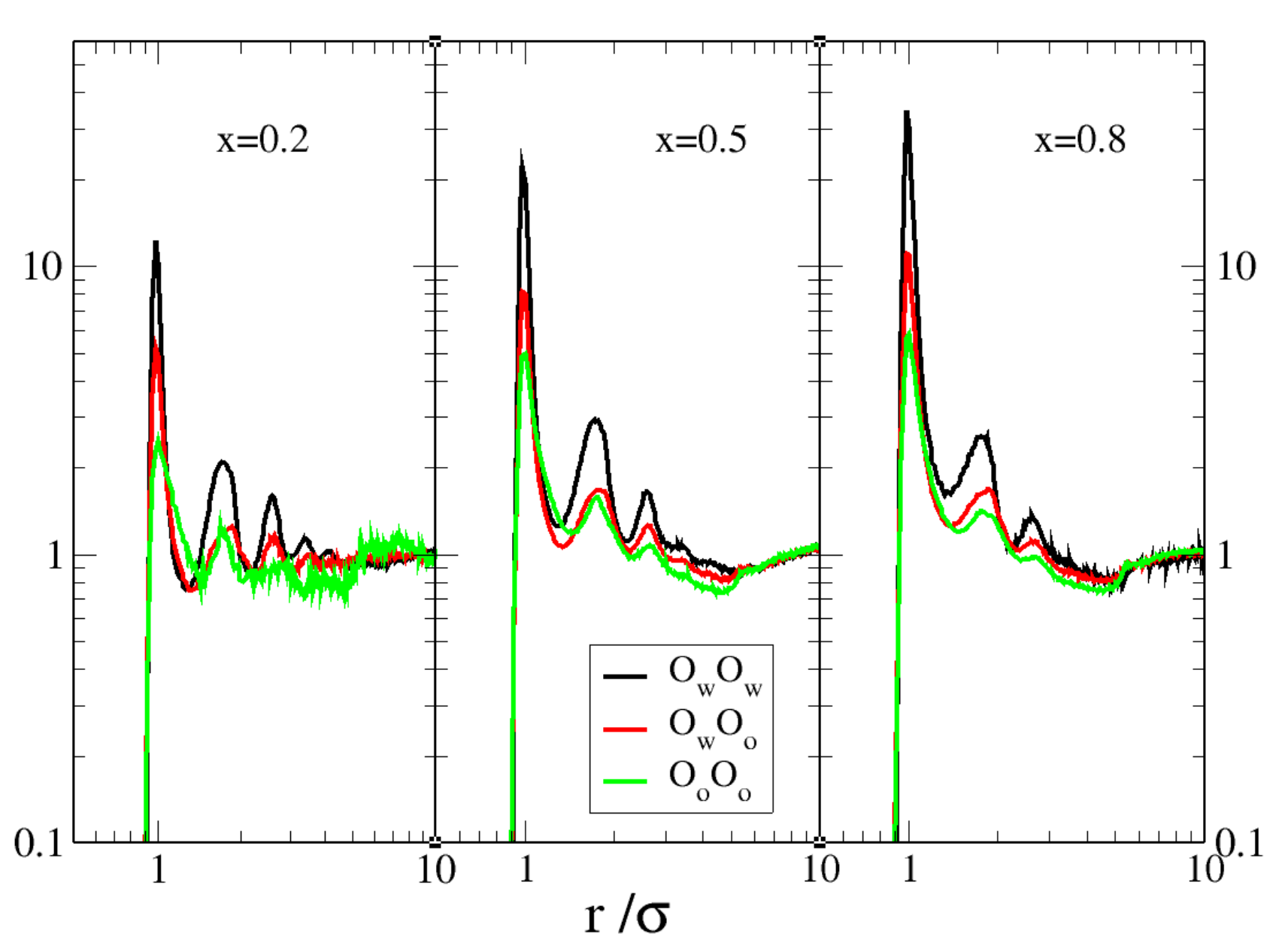}

\caption{Oxygen-oxygen pair correlation functions for three different octanol
concentrations, with same conventions as those in Fig.\ref{Fig=000020GR=000020WM}.
The symbol $O_{o}$ stand for oxygen atoms of \uline{o}ctanol.}

\label{Fig=000020GR=000020WO}
\end{figure}

The snapshots show that, in the case of clear domain formation, most
of the water-alcohol interface is through direct polar head contact.
How does this manifests itself in the polar-head correlation function?
To this intent we examine the various $g_{O_{i}O_{j}}(r)$ site-site
correlation functions between the oxygen atoms. These are shown in
figures \ref{Fig=000020GR=000020WM} , \ref{Fig=000020GR=000020WP}
and \ref{Fig=000020GR=000020WO}, for the aqueous mixtures with methanol,
pentanol and octanol, respectively, and three different concentration.
A quick look shows that they look overall very similar. A second look
reveals that the g(r) for pentanol and octanol are more ``deformed''
than those for methanol, in that they do not converge to 1 in the
usual flat manner. This distortion is the signature of the increased
domain correlations as one goes from methanol to octanol aqueous mixtures.
We now examine the specifics, and separate the discussion on the first
few peaks from that at the long range behaviour.

Fig.\ref{Fig=000020GR=000020WM} show that the $g_{O_{i}O_{j}}(r)$
for the 2D aqueous-methanol mixtures, and for 3 typical alcohol concentration
of 20\%, 50\% and 80\%. As mentioned for Fig.\ref{Fig=000020GR=000020COMP},
water correlations are more structured than the alcohol correlations
and the cross correlations in the entire concentration range. Interestingly,
both the first peak correlations for water and the cross correlations
increase with increasing alcohol concentrations. However, this happens
for different reasons. The reason for the increase of water correlations
is the strenghtening of the water correlations when water is confined
between alcohol domains. 

In contrast, water-alcohols appear to increase with decreasing water
concentration. The reason for cross correlations to increase is because
the water/alcohol polar head contacts happens at the border of the
water domains. Hence, it is a compromise between large contact ``surface''(line)
but few alcohols, and small contact line and more alcohols. We equally
observe a weak increase of the alcohol oxygen correlations, which
is conveniently interpreted as normal expected increase of oxygen
contacts with alcohol content increase. 

Fig.\ref{Fig=000020GR=000020WP} show the $g_{O_{i}O_{j}}(r)$ for
the 2D aqueous pentanol mixtures. These correlations show the same
trends observed for the water-methanol mixtures, except that, for
a given concentration, the first peaks of the water correlations are
systematically higher for pentanol than for methanol, and the same
is somewhat true for the cross correlations. The first peak of the
alcohol correlation also tend to be higher. This can be understood
as an entropic effect due to the longer alkyl tails, which enforce
parallel packing of the alcohols, hence favouring polar head contacts.

Finally, Fig.\ref{Fig=000020GR=000020WO} shows the same trends observed
for water-methanol and water-pentanol. However, we also observe that
the second and higher neighbour peaks of water tend to be systematically
higher. This enforces the picture of globular small water clusters,
confirming again that water is not randomly dispersed in high concentration
alcohol media.

Focusing on the distortions of the shape of the correlations, we explain
them as the superposition of domain correlations onto the atom-atom
correlations. This is not an additive contribution, and appears as
a overall modulation of the atom correlations. This modulation is
nearly invisible in the case of aqueous methanol mixtures, and becomes
progressively more important for pentanol and octanol mixtures, as
could be seen in the snapshots of the previous section. For the case
of octanol, one observes clear domain correlation in the long range
part, as witnessed by the depressing and raising for $r>4\sigma$.
This signature becomes more clear in the study of the structure factors.

\subsection{Structure factors }

\subsubsection{Motivation}

The atom-atom structure factors are defined as:
\[
S_{i_{a}j_{b}}(k)=\delta_{i_{a}j_{b}}+\rho\sqrt{x_{a}x_{b}}\int d\mathbf{r}\exp(i\mathbf{k}.\mathbf{r})\left[g_{a_{i}b_{j}}(r)-1\right]
\]
are related to the Fourier transforms of the site-site pair correlation
functions $g_{a_{i}b_{j}}(r)$, and allow to better visualize the
medium to long range correlations in the system. Indeed, an important
feature is that long range signatures in the $g_{i_{a}j_{b}}(r)$
are transformed into small k features of the $S_{i_{a}j_{b}}(k)$,
and often amplified. This is for instance the case of the weak domain
correlations. In addition, they are related to the scattering intensity
$I(k)$ as measured by radiation scattering techniques\cite{my_monool}.
In the study of pure water and alcohol systems, it was shown that
the $I(k)$ of these model 2D systems were surprisingly similar to
their real 3D counter parts for the case of the x-ray scattering.

\begin{figure}[H]
\centering
\includegraphics[scale=0.2]{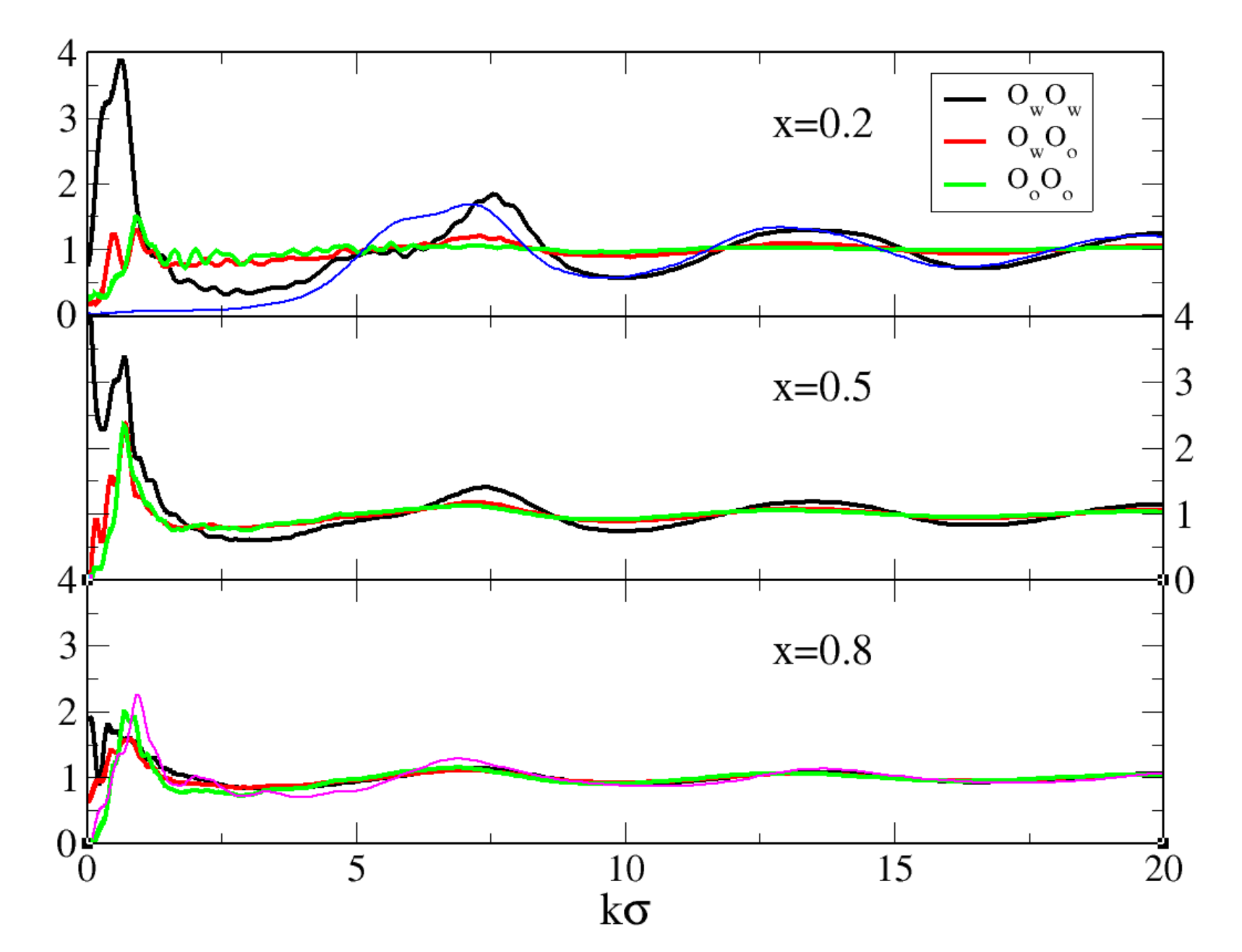}

\caption{Oxygen-oxygen structure factors for three different octanol concentrations:
20\%(upper panel), 50\% (middle panel) and 80\%(lower panel). The
symbols $O_{o}$ stand for oxygen atoms of \uline{p}entanol. The magenta
curve in the lower panel is the structure factor of pure \uline{o}ctanol\cite{AUP-2Dalc}.}

\label{Fig=000020SK=000020WO}
\end{figure}

The structure factors corresponding to the oxygen pair correlations
shown in the subsection above, are displayed in figures \ref{Fig=000020SK=000020WM},
\ref{Fig=000020SK=000020MP} and \ref{Fig=000020SK=000020WO}. Contrary
to the superficial resemblance displayed by the pair correlations,
these functions look very different from each other. This is because
these functions amplify the differences in the long range correlations,
the latter which are not immediately perceptible. We now examine in
the details below.

\subsubsection{Water-methanol}

\begin{figure}[H]
\centering
\includegraphics[scale=0.2]{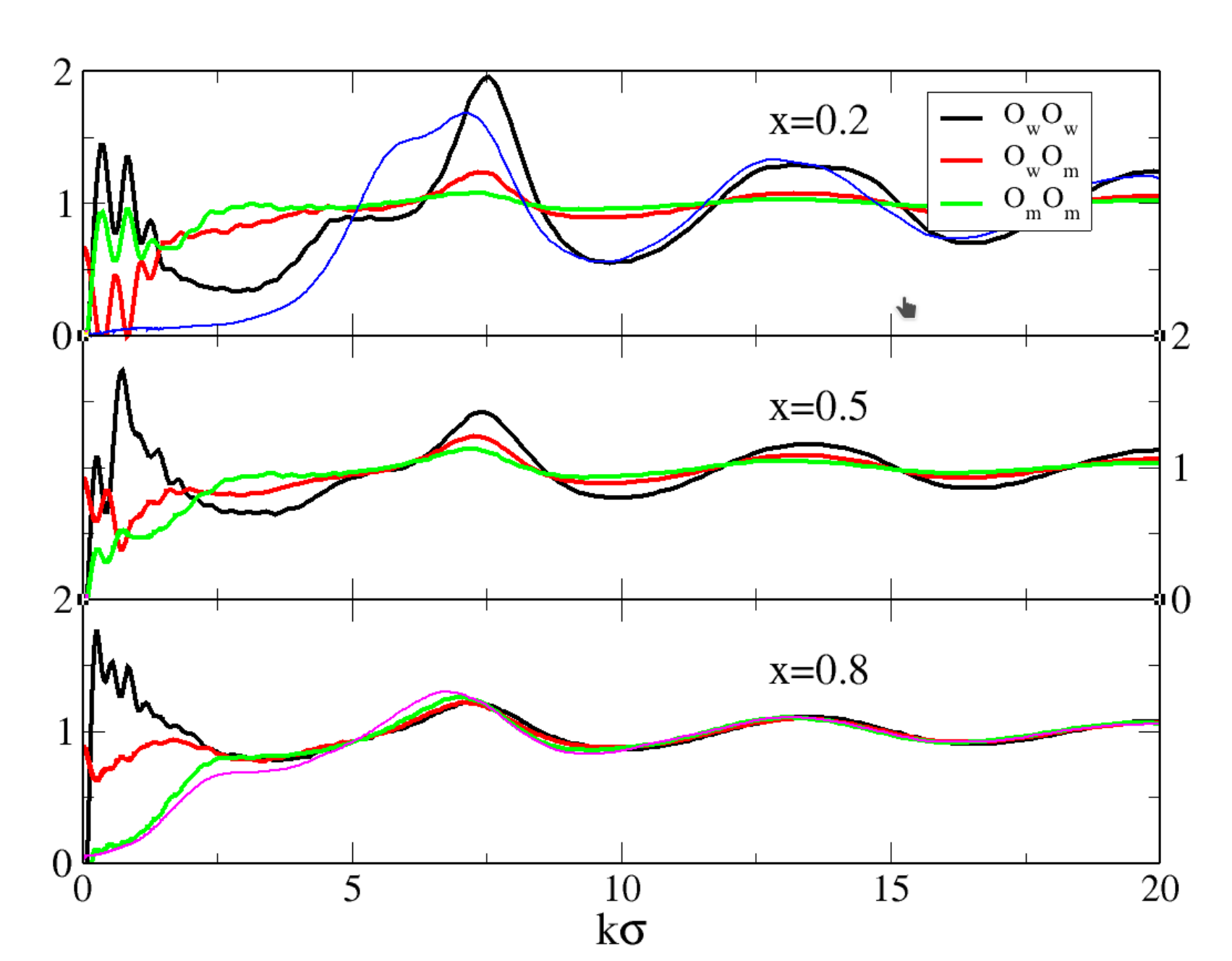}

\caption{Different types of oxygen-oxygen structure factors for three different
methanol concentrations: 20\%(upper panel), 50\% (middle panel) and
80\%(lower panel). The symbols $O_{w}$ and $O_{m}$ stand for oxygen
atoms of \uline{w}ater and \uline{m}ethanol, respectively. The blue
curve in the upper panel is the structure factor for pure water\cite{aupSS-water},
and the magenta curve in the lower panel that of pure methanol\cite{AUP-2Dalc}.}

\label{Fig=000020SK=000020WM}
\end{figure}

Fig.\ref{Fig=000020SK=000020WM} shows the oxygen-oxygen structure
factors $S_{a_{i}b_{j}}(k)$ corresponding to the site-site correlation
functions discussed in the section above. For comparison, the pure
water and pure alcohol structure factors are equally shown in the
upper and lower panels, respectively. Several interesting features
can be observed. For high water content, we see that the characteristic
shoulder shaped first peak, which is a feature of the site model for
water (shown in blue), and which is absent from the original MB water
model\cite{DillMB1998,urbicMB2000}, becomes more spread, with the
inner Hbond shoulder preserved and shifted to higher k ($k\sigma\approx7.5$),
while the lower shoulder is depressed and shifted to lower k values
($k\sigma\approx5$). This indicates that the water-water Hbond correlations
are enforced under mixing conditions, which is compatible with the
observations for confined water in 3D \cite{confinedWater}. In contrast
the direct water-water contacts (second shoulder) are scarcer than
in pure water, but brought to larger distances, since water structure
is quasi exclusively governed by Hbonds. This is fully compatible
with that observed in the snapshots, especially with the Hbonded circular
structures.

The main-peak of pure methanol (lower panel) is depressed and also
shifted to higher k-values ($k\sigma\approx7.5$) indicating tighter
Hbond contacts. In contrast the pre-peak is depressed and shifted
to larger k values ($k\sigma\approx2.5$ instead of $k\sigma\approx3.75$).
This means that the alcohol molecules are less bonded, which is also
what we observe in the snapshots.

The very small k part ($k\sigma<1.5$) shows two behaviour. The first
is an unphysical ringing effect sees by small spurious oscillatory
features, which comes from the 2D Talman transform of the long range
tails in the pair correlation functions, which show small amplitude
oscillations, which correspond to the irregular domain oscillations
of period larger than the molecular separation, hence the small period
of the spurious oscillations. the second feature is the k=0 raise
for water, and to some extent to the two other. These witness the
raise of partial concentration fluctuations, which appear more predominant
for water. We cannot confirm the existence of clear domain pre-peaks,
because of the spurious ringing effects. In any case, such peaks appear
to be hidden by the larger concentration fluctuations.

\subsubsection{Water-pentanol}

\begin{figure}[H]
\centering
\includegraphics[scale=0.2]{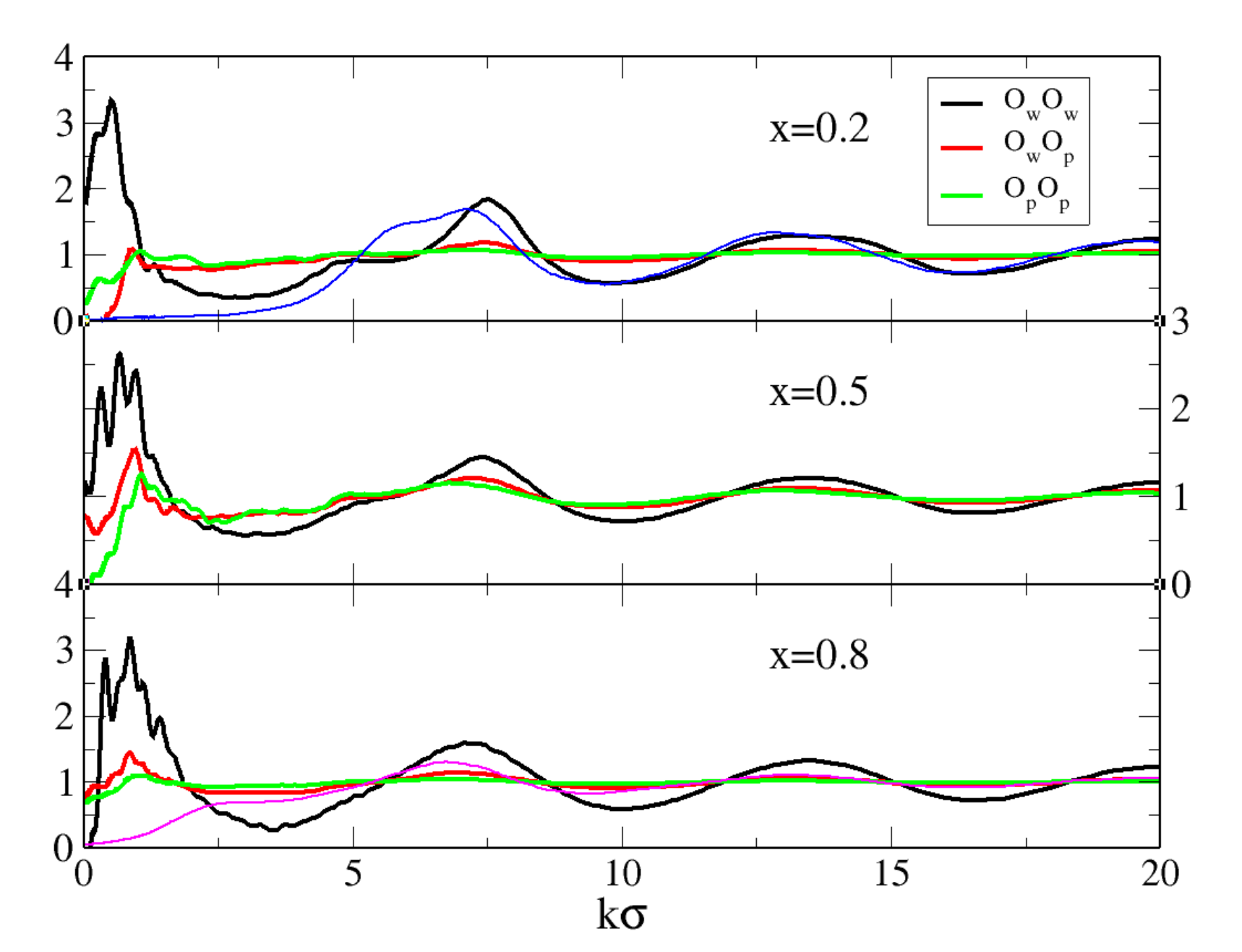}

\caption{Oxygen-oxygen structure factors for three different pentanol concentrations:
20\%(upper panel), 50\% (middle panel) and 80\%(lower panel). The
symbols $O_{p}$ stand for oxygen atoms of \uline{p}entanol. The magenta
curve in the lower panel is the structure factor of pure pentanol\cite{AUP-2Dalc}.}

\label{Fig=000020SK=000020MP}
\end{figure}

Fig.\ref{Fig=000020SK=000020MP} shows the $S_{a_{i}b_{j}}(k)$ for
the pentanol-water mixtures. Globally the features observed are similar
to those seen in the previous case of methanol water. But the concentration
fluctuation effects for water are more pronounced, the $k=0$ raise
bypassing the heights of the main peaks. It suggests enhanced long-wavelength
concentration fluctuations associated with increasingly structured
water-rich domains, as the alcohol tail length increases. This is
understandable since there is more empty room because of the tail
packing problems, as seen from the snapshots in Fig.\ref{snapWP}.

The middle panel for the equimolar mixture appear to suggest a domain
pre-peak for water positioned at $k\sigma\approx1$, which would correspond
to domain structure about 6 diameters (the box size for this system
is about $L\approx60\sigma$). This matches the water aggregates seen
in Fig.\ref{snapWP}. Similar pre-peaks are equally seen for the two
other structure factors and for other concentrations, hence confirming
that the domains are better defined for the alcohol and not so much
affected by fluctuations.

\subsubsection{Water octanol}

Fig.\ref{Fig=000020SK=000020MP} shows the $S_{a_{i}b_{j}}(k)$ for
the octanol-water mixtures. Whereas all the previous observations
still stand for this mixture, for this mixture the water correlations
lead to very clear pre-peak (at $k\sigma\approx1$) for the 20\% alcohol
content (upper panel) and perhaps also for the 50\% content, unfortunately
affected by the ringing effects. But the low water concentration case
show a very clear large concentration fluctuation effect with the
strong raise at $k=0$. The domain formationn for the alcohol is less
pronounced, but present in all the 3 panels. 

The octanol mixtures illustrate a gradual crossover between well-defined
domain pre-peaks and dominant low-$k$ concentration fluctuations
as the composition changes, hence providing a clear illustration of
the paradigm of the duality between the concentration fluctuations
and the micro-heterogeneity with domain formation\cite{myIUPAC,myPCCP-CO,Perera2022,Kolarikova2024}.

\section{Domain correlations and Kirkwood-Buff integrals\protect\label{sec:Domain-correlations-and}}

\subsubsection{Motivation}

The domain correlations appear in the long range part of the $g_{a_{i}b_{j}}(r)$
functions. However, because of the inherent noise as well as the small
magnitude of the oscillations, this is not the most practical way
to look at them. Instead, the running Kirkwood Buff (RKBI) functions
$G_{a_{i}b_{j}}(r)$ allow to visualize smooth domain oscillations,
because of the integrated form
\begin{equation}
G_{a_{i}b_{j}}(r)=4\pi\int_{0}^{r}dss^{2}\left[g_{a_{i}b_{j}}(r)-1\right]\label{RKBI}
\end{equation}
The Kirkwood Buff integrals (KBI) defined as
\begin{equation}
\bar{G}_{a_{i}b_{j}}=\int d\vec{r}\left[g_{a_{i}b_{j}}(r)-1\right]=G_{a_{i}b_{j}}(\infty)\label{KBI}
\end{equation}
can be obtained from the asymptote of the $G_{a_{i}b_{j}}(r)$ , provided
the convergence towards a flat asymptote is achieved at the end of
half box size. For simple mixture this is always the case, provided
the box size is reasonably large. But for mixtures having domain correlations,
the small number of domains, even when the box size is large, does
not allow to reach flat asymptotes. This is illustrated in the following
figures, Fig.\ref{RKBI-MW20}-Fig.\ref{RKBi-WM80} for aqueous methanol
and octanol. The corresponding figures for aqueous pentanol are shown
in the SI document.

\subsubsection{Water methanol}

\begin{figure}[H]
\centering
\includegraphics[scale=0.3]{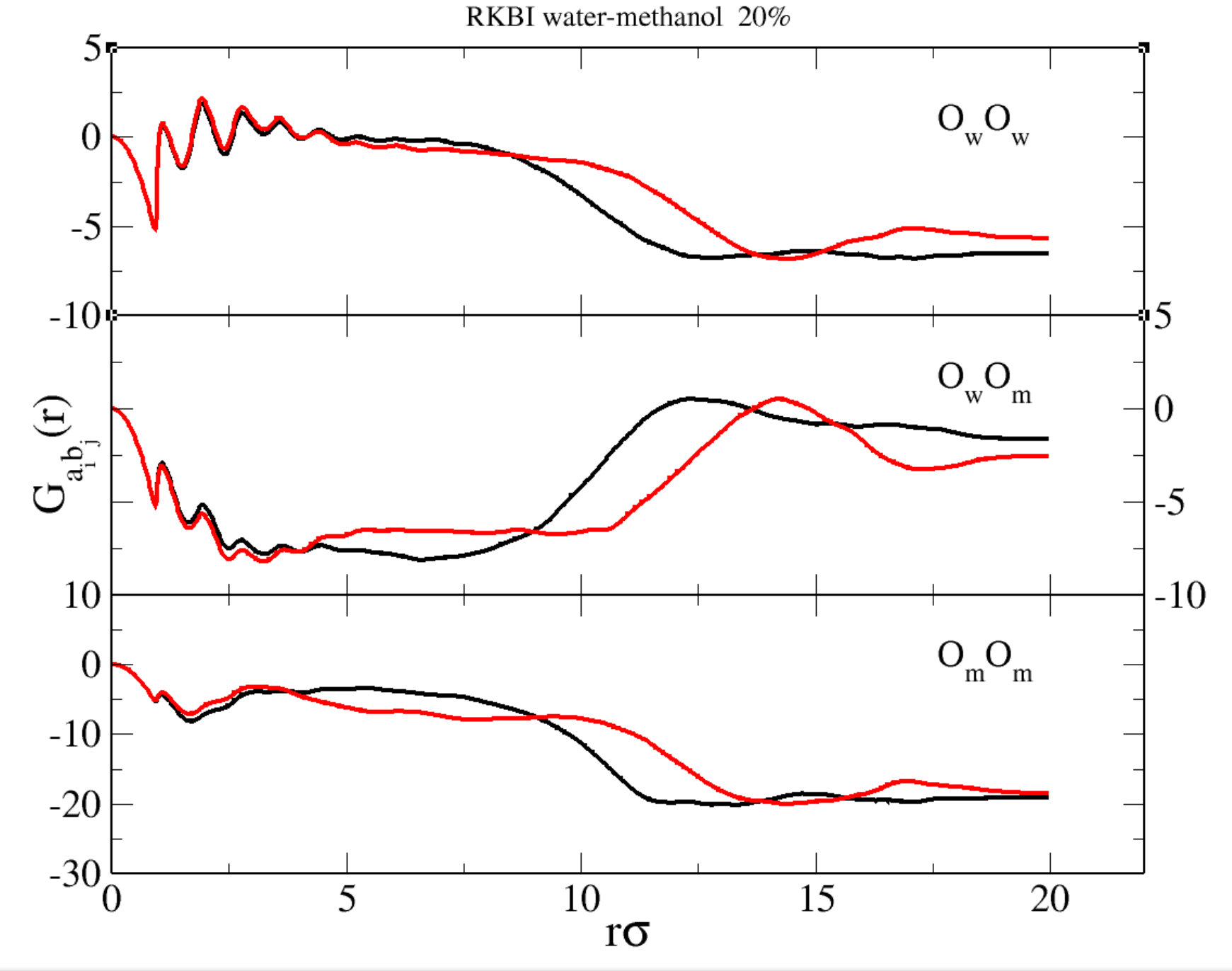}

\caption{Different oxygen-oxygen running Kirkwood-Buff integrals (RKBI) for
the 20\% methanol aqueous mixtures and for two independent statistics
(black and red curves). }

\label{RKBI-MW20}
\end{figure}

The RKBI from the oxygen-oxygen correlations $G_{O_{i}O_{j}}(r)$
of the 2D aqueous-methanol mixtures are shown in Fig.\ref{RKBI-MW20}.
One can see how the atom-atom correlations, witnessed as small oscillations
of width $r/\sigma=1$, are lost at large distances. These oscillations
are more pronounced for the water-water correlations, witnessing the
sustained correlation for water over the cross and alcohol correlations.
The RKBI have been corrected for the asymptote problem (an illustration
of which is provided on few examples in the SI document), but one
can observe two features, one of which is specific to the 20\% aqueous
methanol system and not seen for other concentrations not alcohol.
The generic feature is the apparent absence of self-averaging of the
KBI, illustrated by the fact that two independent runs give different
asymptote, with the important consequence that the exact KBI value
is not precisely determined. We shall discuss this important problem
in the section\ref{sec:Discussion}. 

The second system specific feature is the trend for two asymptotes,
which is clearly visible by the change observed in the range $r/\sigma\approx10-15$.
We interpret this feature as an intermediate stabilisation of the
atom-atom correlations within r/sigma<10, and a second and final domain
influenced stabilisation at larger r-ranges. We tentatively interpret
that this behaviour as a trace of two level RKBI, which becomes visible
for the 20\% aqueous methanol mixtures, but much less for the other
cases, as discussed below. It shows that the asymptotes of the correlations
are stabilized in two ranges, at atomic level for short range, and
domain level for the remaining r-range, and it is the larger one which
is very sensitive to statistics.

Indeed, the domain-domain correlation depend strongly on their sampling.
Within reasonably large system sizes, there is no guaranty that this
problem can be settled. The water domain oscillations are of larger
amplitude for the smaller water domain at 80\% alcohol content. By
contrast, the alcohol has a nice flat asymptote, indicating that the
gr are not affected by domain correlations, since one is closer to
the pure liquid. 

\begin{figure}[H]
\centering
\includegraphics[scale=0.3]{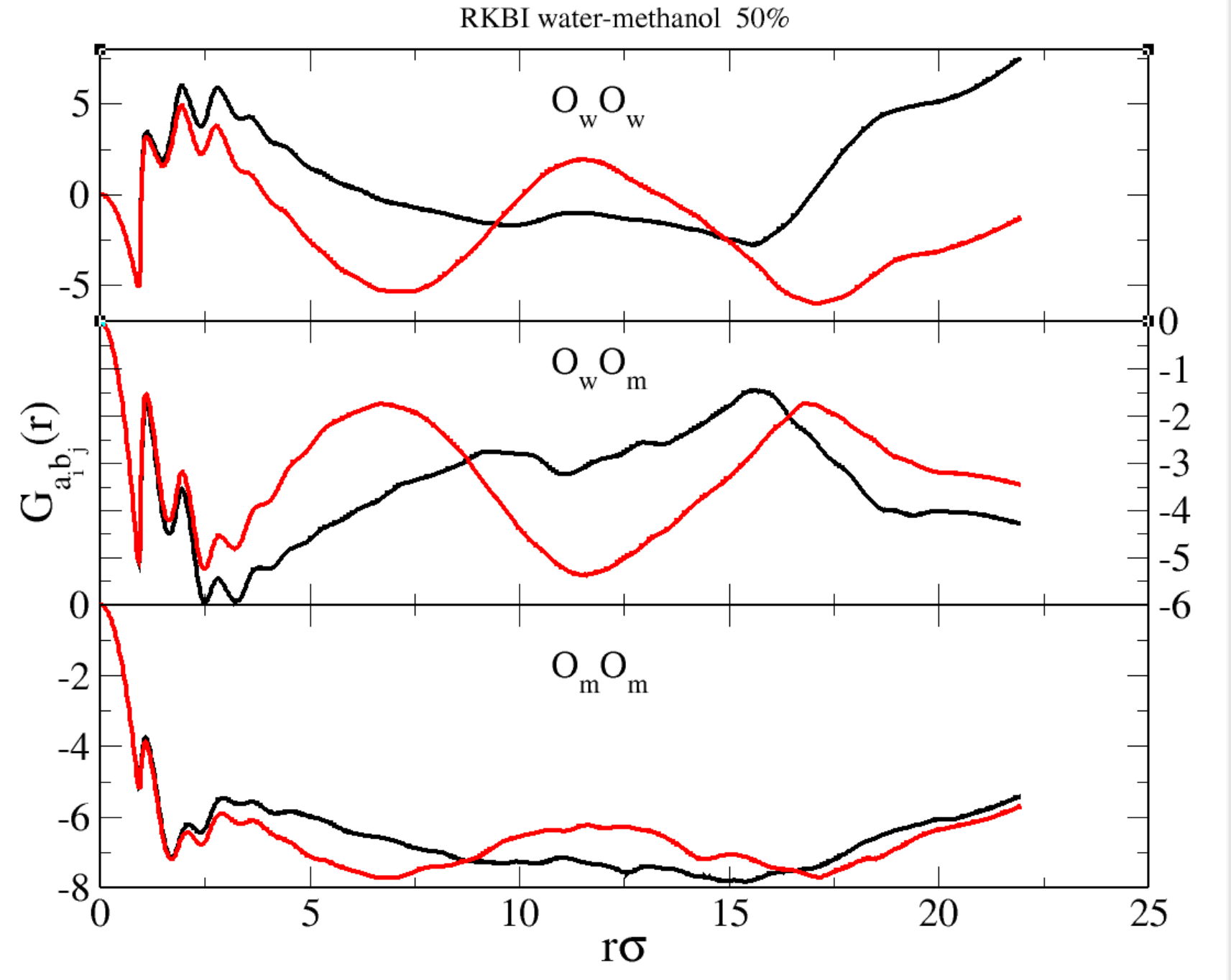}

\caption{Same as in Fig.\ref{RKBI-MW20} but for the 50\% methanol aqueous
mixtures}

\label{RKBI-WM50}
\end{figure}

Fig.\ref{RKBI-WM50} shows the RKBI for the equimolar aqueous methanol
mixtures and for two different runs. In this case, the difference
between the short range and long range is lost and the problem in
self-averaging becomes more apparent. However, the domain oscillations
become very apparent, under the form of oscillations of period $\approx10\sigma$,
more apparent form the water-water correlations that the two others.

The reason for the disappearance of the shorter range RKBI stabilisation
is probably due to more apparent domains, but with fluctuating sizes
and shapes, which does not allow a clear stabilisation of the RKBI
tails.

\begin{figure}[H]
\centering
\includegraphics[scale=0.3]{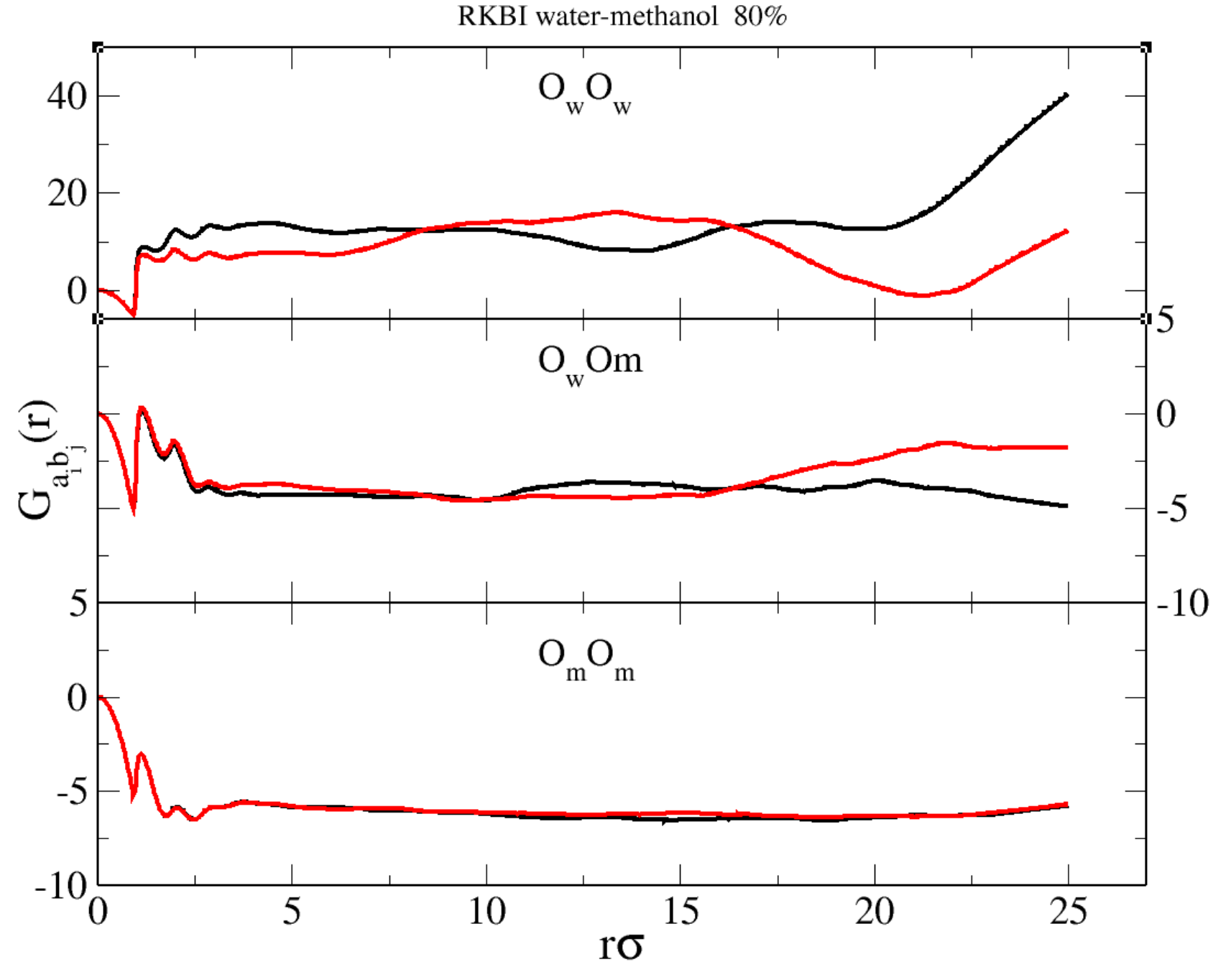}

\caption{Same as in Fig.\ref{RKBI-MW20} but for the 80\% methanol aqueous
mixtures}

\label{RKBi-WM80}
\end{figure}

Fig.\ref{RKBi-WM80} shows the RKBI for the larger alcohol content,
and is is apparent that the asymptote of the RKBI $G_{O_{m}O_{m}}(r)$
is better defined with nearly no domain oscillations, while that of
water $G_{O_{w}O_{w}}(r)$ reflect the underlying domain structure,
with the associated non self-averaging behaviour.

One sees clearly that the simulation box size increases with the alcohol
content because of the packing problem since the latter has a larger
molecule.

\subsubsection{Water octanol}

\begin{figure}[H]
\centering
\includegraphics[scale=0.3]{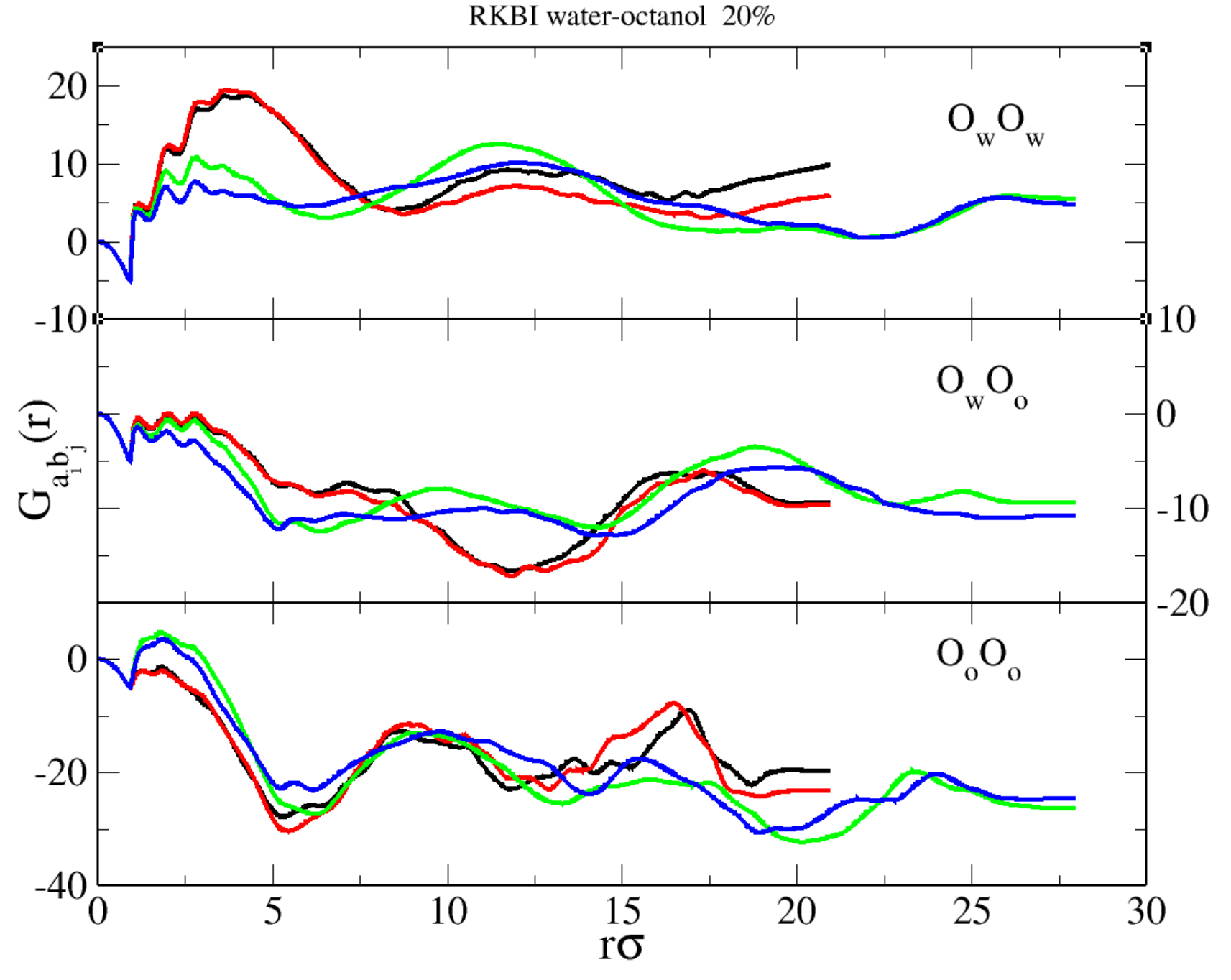}

\caption{Same as in Fig.\ref{RKBI-MW20} but for the 20\% octanol aqueous mixtures.
The black and red curves corresponds to the $N=1000$ particle system
while the green and blue curves to the $N=2000$ particle system.}

\label{RKBi-WO20}
\end{figure}

Fig.\ref{RKBi-WO20} shows the asymptote corrected RKBI for the 20\%
octanol-water mixtures. For this case, we show the RKBI for two different
runs but also two different box sizes, with a larger box with N=2000
particles. The main point is to illustrate the resilient presence
of non-self averaging RKBI tails. In this case, we observe very clearly
the atom-atom oscillations for $r/\sigma<5$ and the domain oscillations
for larger $r/\sigma$ values, particularly for the water\_water correlations.
The fact that the domain correlation modulates the atom-atom correlations
is quite visible in the r/\textbackslash sigma range. 

We also observe that these modulations appear amplified in the smaller
system size N=1000 than in the N=2000 system, particularly so for
the water-water correlations. Indeed, since domain formation is driven
by water, this behaviour is quite understandable.

\begin{figure}[H]
\centering
\includegraphics[scale=0.3]{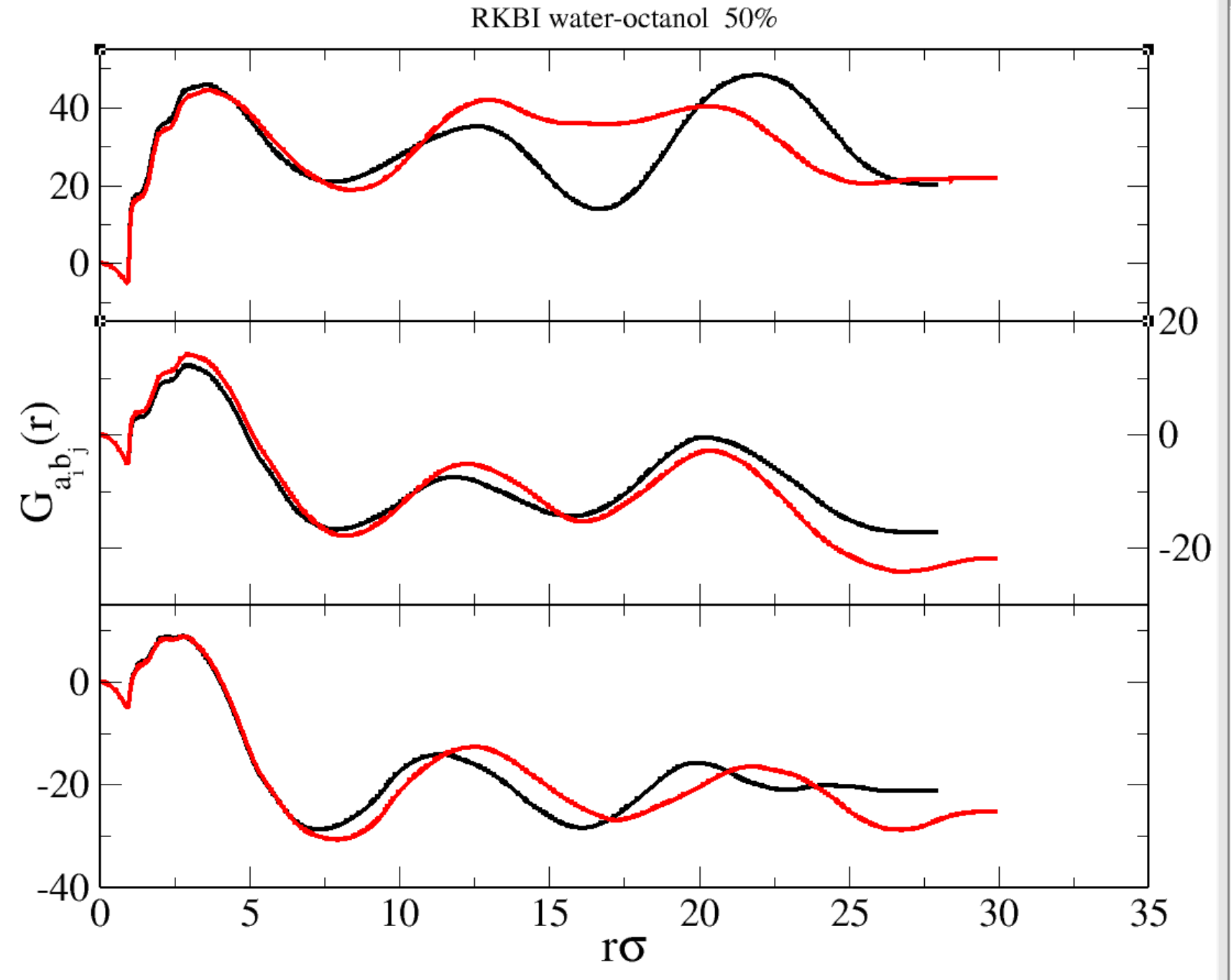}

\caption{Same as in Fig.\ref{RKBI-MW20} but for the 50\% octanol aqueous mixtures}

\label{RKBi-WO50}
\end{figure}

Fig.\ref{RKBi-WO50} shows the asymptote corrected RKBI for the equimolar
aqueous octanol mixtures. The domain correlations are much apparent
and regular than in all the previous cases, but the non-self averaging
remains.

\begin{figure}[H]
\centering
\includegraphics[scale=0.3]{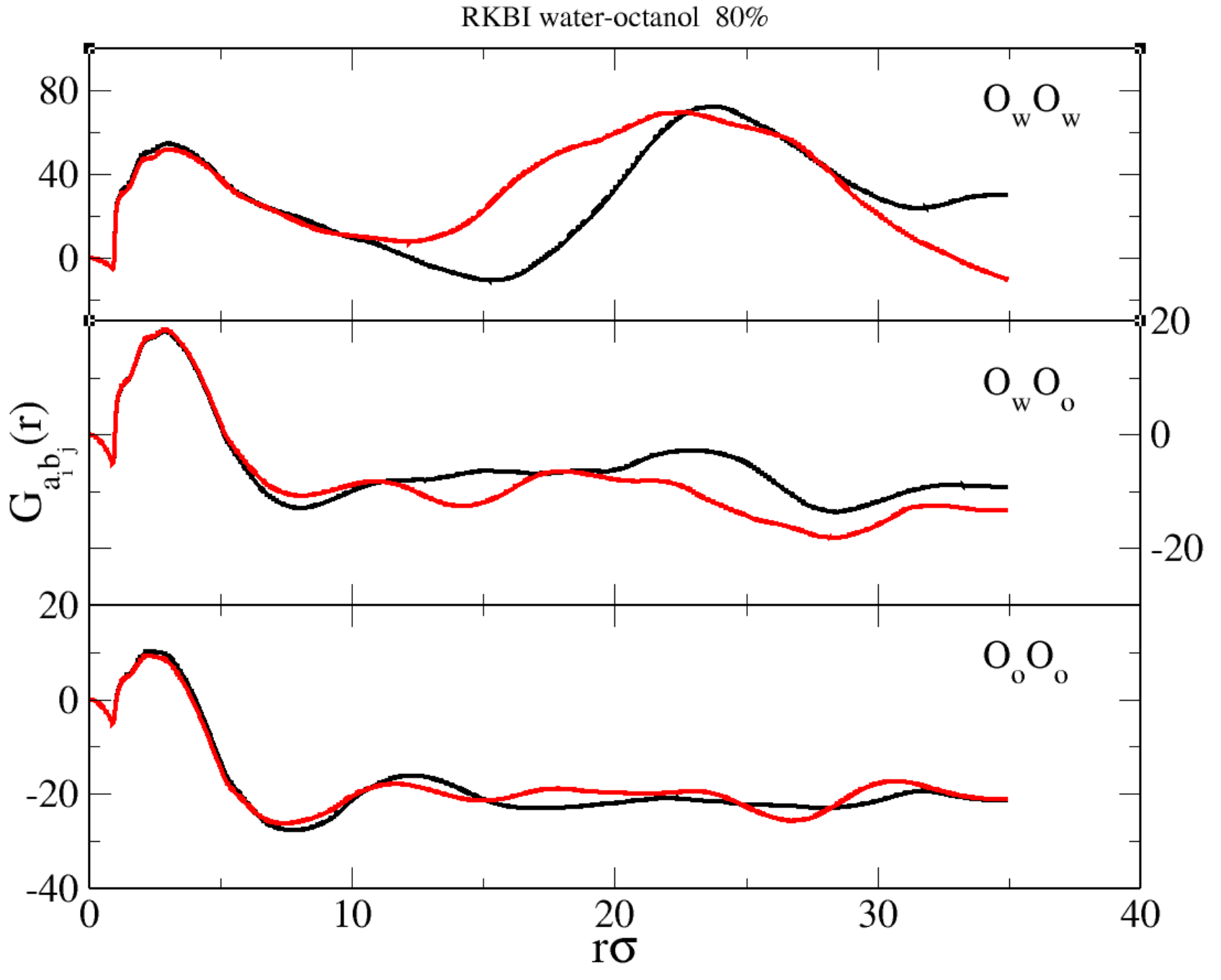}

\caption{}
Same as in Fig.\ref{RKBI-MW20} but for the 80ù octanol aqueous mixtures.

\label{RKBi-WO80}
\end{figure}

Finally, Fig.\ref{RKBi-WO80} shows the RKBI for the 80\% aqueous
octanol mixtures. This time, while one can still observe domain correlations,
we observe that the period for the water-water correlation is larger
than the 50\% case above. This is counter intuitive, since one expects
smaller water domains, namely in view of the right panel in the snapshot
Fig.\ref{snapWO}.

However, what matters in the domain correlation is not the domain
size, but the inter-domain distance, which is indeed greater for the
80\% case than the 50\% case.

We observe that the RKBI for the alcohol converge quite nicely to
a common asymptote between two different runs. This illustrates the
homogeneity of the alcohol distribution, as compared to that water
distribution.

\section{Discussion\protect\label{sec:Discussion}}

The data displayed in this work concerns two different aspects of
these systems. The first one which corresponds to the chemical part
and resides in the differences in the short range part of the pair
correlations $g_{i_{a}j_{b}}(r)$ and the main peak of the corresponding
structure factors $S_{a_{i}b_{j}}(k)$. These parts show strong resemblance
with the 3D systems as displayed in our previous works\cite{myPCCP-CO,myIUPAC}.
The second one resides in the medium-to-long range of the $g_{i_{a}j_{b}}(r)$
and the small-k part of the corresponding $S_{a_{i}b_{j}}(k)$. This
second part is much less discussed in the literature, yet the study
of the 2D systems shows that the it plays an important role in the
analysis of these complex liquids. This was pointed in our previous
works, but the 2D systems clearly amplify the related features. This
tension between the intuitively intepretable short range/medium $k$
features, and the less intuitive long range/small $k$ features, are
the heart of the nature of the fluctuations in these mixtures, and
obviously both in the 3D and 2D. It poses the question whether micro-heterogeneity
is of the same nature as the concentration fluctuations seen near
a demixing mixture\cite{Kolarikova2024}. The 2D analogs of these
mixtures allows to separate these in two manifestations, the usual
fluctuations and the absence of phase separation, transformed into
the pre-peak feature in the structure factors.

The snapshots allow to unmistakably see the domain formation, while
the RKBI show the corresponding domain oscillations. At the same time,
the study of these features points to various problems. Perhaps the
most one is that related to the surprizing absence of self-averaging
in the tail of the RKBI. First of all it is important to note that
this is not a problem related to statistical sampling methods, that
is larger systems and longer runs. At the same time, the asymptotes
of the RKBI are macroscopically well defined quantities, hence one
expects that self-averaging must occur at some system size, possibly
way above the sizes of the present work. Following this line of consideration,
it must be underlined that most of thermo-physical quantities are
well stabilized with system size much less than $N=1000$. 

The idea of simulations to precisely to explore microscopic features
without having to go to macroscopic sizes. Hence, the problem of the
determination of the KBI challenges these implicit criterion. It must
be noted, however, that in all the cases examined above, it was always
possible to bracket the KBI to some extent, hence to provide a mean
estimate of the KBI. This is more challenging in the 3D case, where
the variations of the KBI are much more important. One can provide
an intuitive explanation for the absence of stabilisation of the KBI
over accessible system sizes and times. If the domains are represented
as pseudo-particles with no defined size, shape and contour, then
one faces the problem of two intertwined statistics, the usual particle
statistics and in addition that over the shape of the particle. It
is not obvious that this problem should be solved by merely increasing
the size of the system. A hint to this problem is provided by the
special case of 20\% water-methanol in Fig.\ref{RKBI-MW20}. In this
case, there are two distinct types of aggregate, as witnessed in the
snapshot Fig.\ref{snapMW}: that typical of methanol under the form
of chains, and that typical of water under the quasi hexagonal clusters
and supra clusters. This situation leads to two RKBI regimes. In the
more general case where it is not possible to clearly see this type
of division, it might be an intrinsic feature of the system to exhibit
non-self averaging in some extended intermediate range of $r$ values.
these problems do not occur in the simplistic or less challenging
cases where no micro-heterogeneity exist, such as those examined in
some literature.

It is interesting to compare these 2D aqueous alcohol mixture results
with the recent 2D alcohol-alcohol mixtures\cite{aupAlcMix2D}, where
similar non-self averaging of the RKBI was found. For instance it
was found that alcohol mixtures formed chains of the OH groups. Here,
we do not find that the alcohols have many opportunity to form chain-clusters,
even for the 80\% alcohol concentrations. This is because water tends
to disrupt such formations. In that sense, 2D water acts as a disordering
agent (chaotrope) preventing the alcohol-rich domains from reaching
complete macroscopic separation.

\section{Conclusion}

In this work, a 2D equivalent of aqueous alcohol mixtures was studied,
both for the models themselves in search of common features with their
real 3D analogs, but also to better understand the domain correlations
in relation to the determination of the KBIs through the calculations
of the corresponding RKBIs. Two important features were found. One
hand the 2d aqueous alcohol mixtures remain miscible in all cases,
for all concentrations and long alcohols. This is probably because
the water bonding coherence is restricted by the fluctuations inherent
to the lower dimensionality. The second feature is the domain oscillations
which exhibit non-self averaging, which we conjecture to be a real
feature of such self segregating systems, and occur within a range
which might flirt with mesoscopic size before allowing stable KBI
determination. This problem explains why the determination by computer
simulation of the KBI in micro-heterogeneous 3D real systems poses
so many problems. The present results suggest that micro-segregated
liquids may exhibit an extended intermediate regime where domain correlations
remain insufficiently self-averaged, thereby complicating the determination
of long-range thermodynamic observables. In this respect, 2D models
provide a useful framework to investigate statistical issues that
remain difficult to isolate in realistic 3D systems.

\section*{Acknowledgments}

CdlV and AB thank the Laboratoire de Physique Théorique de la Matière
Condensée for hosting during their L3 internship. 

\bibliographystyle{jpcb_final.bst}
\bibliography{alc2Dmix}

@Article{DillReview,
  author  = {Emiliano Brini and Christopher J. Fennell and Marivi Fernandez - Serra and Barbara HribarLee  and Miha Luk\v{s}i\'{c} and Ken A. Dill},
  journal = {Chemical Reviews},
  title   = {How Water Properties Are Encoded in Its Molecular Structure and Energies},
  year    = {2017},
  note    = {PMID: 28949513},
  number  = {19},
  pages   = {12385-12414},
  volume  = {117},
  doi     = {10.1021/acs.chemrev.7b00259},
  eprint  = {https://doi.org/10.1021/acs.chemrev.7b00259},
  url     = {https://doi.org/10.1021/acs.chemrev.7b00259},
}

@Article{alcTomaz,
  author    = {Papez, Petra and Urbic, Tomaz},
  journal   = {Phys. Rev. E},
  title     = {Simple two-dimensional models of alcohols},
  year      = {2022},
  month     = {May},
  pages     = {054608},
  volume    = {105},
  doi       = {10.1103/PhysRevE.105.054608},
  issue     = {5},
  numpages  = {11},
  publisher = {American Physical Society},
  url       = {https://link.aps.org/doi/10.1103/PhysRevE.105.054608},
}

@Article{BenNaim-MB,
  author  = {Ben - Naim, A.},
  journal = {The Journal of Chemical Physics},
  title   = {{Statistical Mechanics of Waterlike Particles in Two Dimensions. I. Physical Model and Application of the Percus - Yevick Equation}},
  year    = {1971},
  issn    = {0021-9606},
  month   = {05},
  number  = {9},
  pages   = {3682-3695},
  volume  = {54},
  doi     = {10.1063/1.1675414},
  eprint  = {https://pubs.aip.org/aip/jcp/article-pdf/54/9/3682/18873090/3682\_1\_online.pdf},
  url     = {https://doi.org/10.1063/1.1675414},
}

@Article{aupSS-water,
  author    = {Bar\'{e}, Tangi and Besserve, Maxime and Urbic, Tomaz and Perera, Aurélien},
  journal   = {Journal of Molecular Liquids},
  title     = {A site-site interaction two-dimensional model with water like structural properties},
  year      = {2023},
  issn      = {0167-7322},
  month     = sep,
  pages     = {122475},
  volume    = {386},
  doi       = {10.1016/j.molliq.2023.122475},
  publisher = {Elsevier BV},
}

@Article{myIUPAC,
  author    = {Perera, Aurélien},
  journal   = {Pure and Applied Chemistry},
  title     = {From solutions to molecular emulsions},
  year      = {2016},
  issn      = {0033-4545},
  month     = mar,
  number    = {3},
  pages     = {189--206},
  volume    = {88},
  doi       = {10.1515/pac-2015-1201},
  publisher = {Walter de Gruyter GmbH},
}

@Book{chaikin2000principles,
  author    = {Chaikin, P.M. and Lubensky, T.C.},
  publisher = {Cambridge University Press},
  title     = {Principles of Condensed Matter Physics},
  year      = {2000},
  isbn      = {9780521794503},
  lccn      = {93044244},
  url       = {https://books.google.fr/books?id=P9YjNjzr9OIC},
}

@Article{my_monool,
  author    = {Po{\v{z}}ar, Martina and Bolle, Jennifer and Sternemann, Christian and Perera, Aur{\'e}lien},
  journal   = {The Journal of Physical Chemistry B},
  title     = {On the X-ray Scattering Pre-peak of Linear Mono-ols and the Related Microstructure from Computer Simulations},
  year      = {2020},
  issn      = {1520-6106},
  month     = {Sep},
  number    = {38},
  pages     = {8358-8371},
  volume    = {124},
  day       = {24},
  doi       = {10.1021/acs.jpcb.0c05932},
  publisher = {American Chemical Society},
  url       = {https://doi.org/10.1021/acs.jpcb.0c05932},
}

@Article{myPCCP-CO,
  author    = {Perera, Aurélien},
  journal   = {Phys. Chem. Chem. Phys.},
  title     = {Charge ordering and scattering pre-peaks in ionic liquids and alcohols},
  year      = {2017},
  pages     = {1062-1073},
  volume    = {19},
  doi       = {10.1039/C6CP07834F},
  issue     = {2},
  publisher = {The Royal Society of Chemistry},
  url       = {http://dx.doi.org/10.1039/C6CP07834F},
}

@Article{margulis,
  author  = {Hettige, Jeevapani J. and Araque, Juan Carlos and Margulis, Claudio J.},
  journal = {The Journal of Physical Chemistry B},
  title   = {Bicontinuity and Multiple Length Scale Ordering in Triphilic Hydrogen-Bonding Ionic Liquids},
  year    = {2014},
  note    = {PMID: 25157443},
  number  = {44},
  pages   = {12706-12716},
  volume  = {118},
  doi     = {10.1021/jp5068457},
  eprint  = {https://doi.org/10.1021/jp5068457},
  url     = {https://doi.org/10.1021/jp5068457},
}

@Article{SoperNature,
  author   = {Dixit, S. and Crain, J. and Poon, W. C. K. and Finney, J. L. and Soper, A. K.},
  journal  = {Nature},
  title    = {Molecular segregation observed in a concentrated alcohol--water solution},
  year     = {2002},
  issn     = {1476-4687},
  month    = {Apr},
  number   = {6883},
  pages    = {829-832},
  volume   = {416},
  day      = {01},
  doi      = {10.1038/416829a},
  url      = {https://doi.org/10.1038/416829a},
}

@Article{AUP-2Dalc,
  author  = {Aurélien Perera},
  journal = {Journal of Molecular Liquids},
  title   = {Site-site interaction model for alcohol models in two-dimension},
  year    = {2025},
  pages   = {126944},
  volume  = {422},
}

@Article{Perera2022,
  author  = {Perera, Aur\'{e}lien and Po\v{z}ar, Martina and Lovrin\v{c}evi\'{c}, Bernarda},
  journal = {The Journal of Chemical Physics},
  title   = {Camel back shaped Kirkwood - Buff integrals},
  year    = {2022}, 
  issn    = {0021-9606},
  month   = {03},
  number  = {12},
  pages   = {124503},
  volume  = {156},
  doi     = {10.1063/5.0084520},
  eprint  = {https://pubs.aip.org/aip/jcp/article-pdf/doi/10.1063/5.0084520/16540797/124503_1_online.pdf},
  url     = {https://doi.org/10.1063/5.0084520},
}

@Article{Siqueira2011,
  author  = {Siqueira, L.J.A. and Ribeiro, M.C.C.},
  journal = {The Journal of Chemical Physics},
  title   = {Charge ordering and intermediate range order in ammonium ionic liquids},
  year    = {2011},
  number  = {20},
  pages   = {204506},
  volume  = {135},
  doi     = {10.1063/1.3662062},
  eprint  = {https://doi.org/10.1063/1.3662062},
  url     = {https://doi.org/10.1063/1.3662062},
}

@Article{Annapureddy2010,
  author  = {Annapureddy, H.V.R. and Kashyap, H.K. and De Biase, P.M. and Margulis, C.J.},
  journal = {The Journal of Physical Chemistry B},
  title   = {What is the Origin of the Prepeak in the X-ray Scattering of Imidazolium-Based Room-Temperature Ionic Liquids?},
  year    = {2010},
  note    = {PMID: 21077649},
  number  = {50},
  pages   = {16838-16846},
  volume  = {114},
  doi     = {10.1021/jp108545z},
  eprint  = {https://doi.org/10.1021/jp108545z},
  url     = {https://doi.org/10.1021/jp108545z},
}

@Article{Russina2014,
  author  = {Russina, Olga and Sferrazza, Alessio and Caminiti, Ruggero and Triolo, Alessandro},
  journal = {The Journal of Physical Chemistry Letters},
  title   = {Amphiphile Meets Amphiphile: Beyond the Polar - Apolar Dualism in Ionic Liquid/Alcohol Mixtures},
  year    = {2014},
  note    = {PMID: 26270376},
  number  = {10},
  pages   = {1738-1742},
  volume  = {5},
  doi     = {10.1021/jz500743v},
  eprint  = {https://doi.org/10.1021/jz500743v},
  url     = {https://doi.org/10.1021/jz500743v},
}

@Article{Sedlak2014,
  author  = {Sedl\'{a}k, Mari\'{a}n and Rak, Dmytro},
  journal = {The Journal of Physical Chemistry B},
  title   = {On the Origin of Mesoscale Structures in Aqueous Solutions of Tertiary Butyl Alcohol: The Mystery Resolved},
  year    = {2014},
  note    = {PMID: 24559045},
  number  = {10},
  pages   = {2726-2737},
  volume  = {118},
  doi     = {10.1021/jp500953m},
  eprint  = {https://doi.org/10.1021/jp500953m},
  url     = {https://doi.org/10.1021/jp500953m},
}

@Article{Kolarikova2024,
  author  = {Kola\v{r}\'{i}kov\'{a}, Alena and Perera, Aurélien},
  journal = {Journal of Chemical Theory and Computation},
  title   = {Concentration Fluctuation/Microheterogeneity Duality Illustrated with Aqueous 1,4-Dioxane Mixtures},
  year    = {2024},
  note    = {PMID: 38687823},
  number  = {9},
  pages   = {3473-3483},
  volume  = {20},
  doi     = {10.1021/acs.jctc.4c00151},
  eprint  = {https://doi.org/10.1021/acs.jctc.4c00151},
  url     = {https://doi.org/10.1021/acs.jctc.4c00151},
}

@Article{Friedrich2025,
  author  = {Friedrich, Lena and Po\v{z}ar, Martina and Perera, Aur\'{e}lien and Paulus, Michael and Thiering, Nicola and Savelkous, Jaqueline and Schneider, Eric and Lovrin\'{c}evi\'{c} and L\"{u}tzenkirchen-Hect, Dirk and Sternemann, Christian},
  journal = {currently in review},
  title   = {Molecular micro-heterogeneity: Structure formation in amine-water mixtures},
  year    = {2025},
}

@Article{Overduin2019,
  author   = {Overduin, S. D. and Perera, Aur\'{e}lien and Patey, G. N.},
  journal  = {The Journal of Chemical Physics},
  title    = {{Structural behavior of aqueous t-butanol solutions from large-scale molecular dynamics simulations}},
  year     = {2019},
  issn     = {0021-9606},
  month    = may,
  note     = {184504},
  number   = {18},
  volume   = {150},
  doi      = {10.1063/1.5097011},
  eprint   = {https://pubs.aip.org/aip/jcp/article-pdf/doi/10.1063/1.5097011/13944266/184504\_1\_online.pdf},
  url      = {https://doi.org/10.1063/1.5097011},
}

@Article{Overduin2017,
  author   = {Overduin, S. D. and Patey, G. N.},
  journal  = {The Journal of Chemical Physics},
  title    = {{Comparison of simulation and experimental results for a model aqueous tert-butanol solution}},
  year     = {2017},
  issn     = {0021-9606},
  month    = jul,
  note     = {024503},
  number   = {2},
  volume   = {147},
  doi      = {10.1063/1.4990505},
  eprint   = {https://pubs.aip.org/aip/jcp/article-pdf/doi/10.1063/1.4990505/13978255/024503\_1\_online.pdf},
  url      = {https://doi.org/10.1063/1.4990505},
}

@Article{Triolo2007,
  author    = {Triolo, Alessandro and Russina, Olga and Bleif, Hans-Jurgen and Di Cola, Emanuela},
  journal   = {J. Phys. Chem. B},
  title     = {Nanoscale Segregation in Room Temperature Ionic Liquids},
  year      = {2007},
  issn      = {1520-6106},
  month     = may,
  number    = {18},
  pages     = {4641--4644},
  volume    = {111},
  comment   = {doi: 10.1021/jp067705t},
  doi       = {10.1021/jp067705t},
  publisher = {American Chemical Society},
  url       = {https://doi.org/10.1021/jp067705t},
}

@Article{Brovchenko2001,
  author   = {I. Brovchenko and B. Guillot},
  journal  = {Fluid Phase Equilibria},
  title    = {Simulation of the liquid-liquid coexistence of the tetrahydrofuran+water mixture in the Gibbs ensemble},
  year     = {2001},
  issn     = {0378-3812},
  note     = {Proceedings of the fourteenth symposium on thermophysical properties},
  pages    = {311-319},
  volume   = {183-184},
  doi      = {https://doi.org/10.1016/S0378-3812(01)00443-5},
  url      = {https://www.sciencedirect.com/science/article/pii/S0378381201004435},
}

@Article{THF-problem,
  author  = {N. F. Bunkin and A. V. Shkirin and G. A. Lyakhov and A. V. Kobelev and N. V. Penkovi and S. V. Ugraitskaya and E. E. Fesenko},
  journal = {The Journal of Chemical Physics},
  title   = {Droplet-like heterogeneity of aqueous tetrahydrofuran solutions at thesubmicrometer scale},
  year    = {2016},
  pages   = {184501},
  volume  = {145},
  doi     = {10.1063/1.4966187},
}

@Article{smith-TFE,
  author  = {Rajappa Chitra and Paul E. Smith},
  journal = {The Journal of Chemical Physics},
  title   = {Properties of 2,2,2-trifluoroethanol and water mixtures},
  year    = {2001},
  pages   = {426},
  volume  = {114},
}

@Article{nico-TBA,
  author  = {Maeng Eun Lee and Nico F. A. van der Vegt},
  journal = {The Journal of Chemical Physics},
  title   = {A new force field for atomistic simulations of aqueoustertiary butanol solutions},
  year    = {2005},
  pages   = {114509},
  volume  = {122},
  doi     = {10.1063/1.1862625},
}

@Article{aup-POF,
  author  = {Aur\'{e}lien Perera},
  journal = {Physics of fluids},
  title   = {Mesoscopic correlations in aqueous alkylaminemixtures between molecular and micro emulsions},
  year    = {2026},
  pages   = {017125},
  volume  = {38},
  doi     = {10.1063/5.0311277},
}

@Article{self-av-WisemanDomany1995,
  author  = {Wiseman, Itamar and Domany, Eytan},
  journal = {Physical Review Letters},
  title   = {Lack of Self-Averaging in Critical Disordered Systems},
  year    = {1998},
  number  = {1},
  pages   = {22--25},
  volume  = {81},
  doi     = {10.1103/PhysRevLett.81.22},
}

@Article{self-av-AharonyHarris1996,
  author  = {Aharony, Aharon and Harris, A. B.},
  journal = {Physical Review Letters},
  title   = {Absence of Self-Averaging and Universal Fluctuations in Random Systems Near Critical Points},
  year    = {1996},
  number  = {18},
  pages   = {3700--3703},
  volume  = {77},
  doi     = {10.1103/PhysRevLett.77.3700},
}

@Book{self-av-BinderStauffer1997,
  author    = {Binder, Kurt and Stauffer, Dietrich},
  publisher = {Cambridge University Press},
  title     = {A Guide to Monte Carlo Simulations in Statistical Physics},
  year      = {1997},
  doi       = {10.1017/CBO9780511613487},
}

@Article{anisimov2011,
  author       = {Subramanian, Deepa and Ivanov, Dmitry A. and Yudin, Igor K. and Anisimov, Mikhail A. and Sengers, Jan V.},
  journal      = {Journal of Chemical and Engineering Data},
  title        = {Mesoscale {{Inhomogeneities}} in {{Aqueous Solutions}} of 3-{{Methylpyridine}} and {{Tertiary Butyl Alcohol}}},
  year         = {2011},
  issn         = {0021-9568, 1520-5134},
  number       = {4},
  pages        = {1238--1248},
  volume       = {56},
  date         = {2011-04-14},
  doi          = {10.1021/je101125e},
  journaltitle = {Journal of Chemical \& Engineering Data},
  shortjournal = {J. Chem. Eng. Data},
  url          = {https://pubs.acs.org/doi/10.1021/je101125e},
  urldate      = {2026-06-02},
}

@Article{aup_faraday,
  author       = {Perera, Aur\'{e}lien and Ke\v{z}i\`{c}, Bernarda},
  journal      = {Faraday discussions},
  title        = {Fluctuations and Micro-Heterogeneity in Mixtures of Complex Liquids},
  year         = {2013},
  issn         = {1359-6640, 1364-5498},
  pages        = {145},
  volume       = {167},
  date         = {2013},
  doi          = {10.1039/c3fd00072a},
  journaltitle = {Faraday Discussions},
  langid       = {english},
  shortjournal = {Faraday Discuss.},
  url          = {https://xlink.rsc.org/?DOI=c3fd00072a},
  urldate      = {2026-06-02},
}

@Article{kezic-Coresoft-2015,
  author       = {Ke\v{z}i\`{c}-Lovrin\v{c}evi\`{c}, Bernarda and Dartois, St\'{e}phane and Perera, Aur\'{e}lien},
  journal      = {Molecular Physics},
  title        = {Repulsive Core-Soft Models for Binary Aqueous Mixtures},
  year         = {2015},
  issn         = {0026-8976, 1362-3028},
  number       = {9--10},
  pages        = {1108--1118},
  volume       = {113},
  date         = {2015-05-19},
  doi          = {10.1080/00268976.2015.1005189},
  journaltitle = {Molecular Physics},
  langid       = {english},
  shortjournal = {Molecular Physics},
  url          = {http://www.tandfonline.com/doi/full/10.1080/00268976.2015.1005189},
  urldate      = {2026-06-02},
}

@Article{aup_baptista,
  author       = {Baptista, Anthony and Perera, Aur\'{e}lien},
  journal      = {the Journal of Chemical Physics},
  title        = {Modeling Micro-Heterogeneity in Mixtures: {{The}} Role of Many Body Correlations},
  year         = {2019},
  issn         = {0021-9606, 1089-7690},
  number       = {6},
  pages        = {064504},
  volume       = {150},
  date         = {2019-02-14},
  doi          = {10.1063/1.5066598},
  journaltitle = {The Journal of Chemical Physics},
  shorttitle   = {Modeling Micro-Heterogeneity in Mixtures},
  url          = {https://pubs.aip.org/jcp/article/150/6/064504/199042/Modeling-micro-heterogeneity-in-mixtures-The-role},
  urldate      = {2026-06-02},
}

@Article{urbicMB2000,
  author       = {Urbi\v{c}, T. and Vlachy, V. and Kalyuzhnyi, Yu. V. and Southall, N. T. and Dill, K. A.},
  journal      = {The Journal of Chemical Physics},
  title        = {A Two-Dimensional Model of Water: {{Theory}} and Computer Simulations},
  year         = {2000},
  issn         = {0021-9606, 1089-7690},
  number       = {6},
  pages        = {2843--2848},
  volume       = {112},
  date         = {2000-02-08},
  doi          = {10.1063/1.480928},
  journaltitle = {The Journal of Chemical Physics},
  shorttitle   = {A Two-Dimensional Model of Water},
  url          = {https://pubs.aip.org/jcp/article/112/6/2843/473768/A-two-dimensional-model-of-water-Theory-and},
  urldate      = {2026-06-02},
}

@Article{DillMB1998,
  author       = {Silverstein, Kevin A. T. and Haymet, A. D. J. and Dill, Ken A.},
  journal      = {Journal of the American Chemical Society},
  title        = {A {{Simple Model}} of {{Water}} and the {{Hydrophobic Effect}}},
  year         = {1998},
  issn         = {0002-7863, 1520-5126},
  number       = {13},
  pages        = {3166--3175},
  volume       = {120},
  date         = {1998-04-01},
  doi          = {10.1021/ja973029k},
  journaltitle = {Journal of the American Chemical Society},
  langid       = {english},
  shortjournal = {J. Am. Chem. Soc.},
  url          = {https://pubs.acs.org/doi/10.1021/ja973029k},
  urldate      = {2026-06-02},
}

@Article{urbicMB2002,
  author       = {Urbi\v{c}, T. and Vlachy, V. and Kalyuzhnyi, Yu. V., V and Southall, N. T. and Dill, K. A.},
  journal      = {The Journal of Chemical Physics},
  title        = {A Two-Dimensional Model of Water: {{Solvation}} of Nonpolar Solutes},
  year         = {2002},
  issn         = {0021-9606, 1089-7690},
  number       = {2},
  pages        = {723--729},
  volume       = {116},
  date         = {2002-01-08},
  doi          = {10.1063/1.1427307},
  journaltitle = {The Journal of Chemical Physics},
  shorttitle   = {A Two-Dimensional Model of Water},
  url          = {https://pubs.aip.org/jcp/article/116/2/723/442540/A-two-dimensional-model-of-water-Solvation-of},
  urldate      = {2026-06-02},
}

@Article{AqOct-siepmann,
  author  = {Bin Chen and J. Ilja Siepmann},
  journal = {J. Phys. Chem. B},
  title   = {Microscopic Structure and Solvation in Dry and Wet Octanol},
  year    = {2006},
  pages   = {3555 - 3563},
  volume  = {110},
  doi     = {10.1021/jp0548164},
}

@Article{confinedWater,
  author  = {Marie-Claire Bellissent-Funel},
  journal = {Journal of Molecular Liquids},
  title   = {Structure and dynamics of confined water: Selected examples},
  year    = {2023},
  pages   = {123370},
  volume  = {391},
  doi     = {10.1016/j.molliq.2023.123370},
}

@Online{aupAlcMix2D,
  author      = {Chelli, Lydia and Perera, Aurélien},
  date        = {2026-04-24},
  doi         = {10.48550/arXiv.2604.22936},
  eprint      = {2604.22936},
  eprintclass = {physics.chem-ph},
  eprinttype  = {arXiv},
  title       = {Charge Order, Domain Order, Ideal Mixing and Absence of Demixing in {{2D}} Binary Mixtures of Alcohols},
  url         = {http://arxiv.org/abs/2604.22936},
  urldate     = {2026-06-03},
}

\end{document}